\documentclass[10pt]{article}
\usepackage{bbm}
\usepackage[dvips]{graphicx}
\usepackage{amsmath,amsfonts,amsbsy,amssymb}
\usepackage{latexsym}
\usepackage[latin1]{inputenc}

\def\empile#1\over#2{\mathrel{\mathop{\kern 0pt#1}\limits_{#2}}}

\newcommand{\nr}[1]{(\ref{#1})} 

\newcommand{\fig}{fig.~}
\newcommand{\figs}{figs.~}
\newcommand{\eq}{eq.~}
\newcommand{\se}{sec.~}
\newcommand{\eqs}{eqs.~}
\def\p{{\boldsymbol p}}
\def\q{{\boldsymbol q}}

\def\k{{\boldsymbol k}}

\def\x{{\boldsymbol x}}
\def\y{{\boldsymbol y}}

\newcommand{\bE}{\boldsymbol{E}}
\newcommand{\bA}{\boldsymbol{A}}
\newcommand{\rmd}{\mathrm{d}}
\newcommand{\rmi}{\mathrm{i}}
\newcommand{\rme}{\mathrm{e}}
\newcommand{\Hyg}[4]{{}_2F_1(#1,#2\,;\,#3\,;\,#4)}
\newcommand{\half}{{\scriptstyle \frac{1}{2}}}

\newcommand{\rmout}{\textrm{out},}
\newcommand{\rmin}{\textrm{in},}

\def\bs{\boldsymbol}

\begin{document}

\title{\bf Multiparticle correlations\\ in the Schwinger mechanism}

\date{}

\author{Kenji Fukushima$^{(1)}$, Fran\c cois Gelis$^{(2)}$,
        Tuomas Lappi$^{(2,3)}$}
\maketitle
\begin{center}
\begin{enumerate}
\item Yukawa Institute for Theoretical Physics, Kyoto University, \\
  Oiwake-cho, Kitashirakawa, Sakyo-ku, \\
  Kyoto 606-8502, Japan
\item Institut de Physique Th\'eorique (URA 2306 du CNRS)\\
  CEA/DSM/Saclay, B\^at.\ 774\\
  91191, Gif-sur-Yvette Cedex, France
\item  Department of Physics, \\
 P.O. Box 35, 40014 University of Jyv\"askyl\"a, Finland
\end{enumerate}
\end{center}

\maketitle

\begin{abstract}
  We discuss the Schwinger mechanism in scalar QED and derive the
  multiplicity distribution of particles created under an
  external electric field using the LSZ reduction formula.  Assuming
  that the electric field is spatially homogeneous, we find that the
  particles of different momenta are produced independently, and that
  the multiplicity distribution in one mode follows a Bose-Einstein
  distribution.  We confirm the consistency of our results with an
  intuitive derivation by means of the Bogoliubov transformation on
  creation and annihilation operators.  Finally we revisit a known
  solvable example of time-dependent electric fields to present exact
  and explicit expressions for demonstration.
\end{abstract}

\begin{flushright}
{\small\sf YITP-09-45}
\\
{\small\sf IPhT-t09/104}
\end{flushright}

\section{Introduction}
A classic example of a non-perturbative tunneling phenomenon in
quantum field theory is the decay of an electric field due to pair
creation.  The quantum vacuum is full of virtual particle-antiparticle
pairs (i.e.\ the Dirac sea), which can occasionally gain enough energy
from the external field to become real.  The decay (or persistence)
rate of the QED (quantum electrodynamics) vacuum in the presence of an
external electric field was first deduced from the imaginary part of
the Heisenberg-Euler Lagrangian~\cite{Heisenberg:1935qt} and
formulated in Schwinger's classic paper~\cite{Schwinger:1951nm}. The
phenomenon is commonly referred to as the \textit{Schwinger mechanism}
(see ref.~\cite{Dunne:2004nc} for a comprehensive review.)
In the case of QED the coupling constant $e$ is very small, and it is
difficult in practice to achieve large enough electric fields; the
probability of producing an electron-positron pair is, up to a
prefactor, $\sim\exp[-\pi m_e^2/(eE)]$, and thus a very strong
electric field, $E \sim m_e^2 /e\simeq 1.3\times 10^{18}\;\mbox{V/m}$,
is necessary to observe the phenomenon.  To the best of our knowledge,
the Schwinger mechanism in QED remains to be unambiguously
observed experimentally.

Pair creation from an electric field became a phenomenologically much
more relevant subject with the realization that the strong nuclear
force is described by a gauge theory called QCD (quantum
chromo-dynamics).  A popular phenomenological view of QCD with
confinement is a description in terms of a chromoelectric flux tube
connecting the color charges of the quarks.  If these quarks are then
pulled apart by their momenta, the string formed by the chromoelectric
field can decay via the Schwinger mechanism leading to the decay of
the system into $q\bar{q}$ or color neutral mesons as a result of
hadronization.  In this case, the decay probability is characterized
by $\sim\exp[-\pi m_q^2/\sigma]$, where the QCD string tension
$\sigma\simeq 1\;\text{GeV}/\text{fm}$ is an energy stored in the
chromoelectric flux tube per unit length.  Applications of the
particle production by the Schwinger mechanism range from $e^+ e^-$
annihilation~\cite{Andersson:1983ia,Bialas:1989hc} to early models of
relativistic heavy ion
collisions~\cite{Casher:1978wy,Glendenning:1983qq,Biro:1984cf,%
  Bialas:1984ye,Kajantie:1985jh,Gatoff:1987uf,Wang:1988ct}.  Recent
extensive studies on thermal hadron production as a possible
manifestation of the Hawking-Unruh effect, that is an equivalent
formulation to the Schwinger mechanism in curved space-time, is found
in refs.~\cite{Kharzeev:2005iz,Kharzeev:2006zm,Castorina:2007eb}.

The QCD coupling constant $g$, although asymptotically small, is not
as small as the QED one at phenomenologically interesting energies.
Even more important is that the nonlinear dynamics of the gauge fields
naturally leads, in some circumstances, to gauge fields that are
parametrically large in the coupling,  $A_\mu \sim 1/g$.  A
prime example of such a situation is caused by the large occupation
numbers of gluonic states in high energy scattering.  The
transverse gluon density $\sim Q_\mathrm{s}^2$ provides a typical energy scale
$Q_\mathrm{s}$ in such a system.  There the nonlinear interactions among
bremsstrahlung gluons with small Bjorken's $x$ lead to gluon
saturation, which is most conveniently described as a coherent color
field radiated by static (in light cone time) sources.  This
description is referred to as the Color Glass Condensate (CGC) (for
reviews,
see~\cite{Iancu:2002xk,Iancu:2003xm,Weigert:2005us,Gelis:2007kn}).
The collision of two objects whose wavefunction is characterized by
$Q_s$ in the CGC formalism achieves a field configuration of
longitudinal chromoelectric and chromomagnetic fields whose strength
is also given by $Q_s$.  This transient state of matter containing
strong longitudinal fields is known as the
\textit{glasma}~\cite{Kovner:1995ja,Kovner:1995ts,Lappi:2006fp}.  In
the case of QCD it is of course difficult to achieve the canonical
model case of a constant electrical field.  Generically, a WKB-type
non-perturbative evaluation such as in
refs.~\cite{Kharzeev:2005iz,Kharzeev:2006zm} could be expected to be
valid in a case where the field strength $gA_\mu\sim Q_s$ is much
larger than the typical (inverse) time and spatial scales of the
fields.  A perturbative calculation, on the other hand, is also
feasible in the case when $g$ is small enough.  The particle (gluons
and quarks) production associated with strong CGC fields has been
formulated based on the Lehmann-Symanzik-Zimmermann (LSZ) reduction
formula~\cite{Blaizot:2004wu,Blaizot:2004wv,Baier:2005dv,%
  Gelis:2006yv,Fujii:2006ab,Gelis:2006cr,Marquet:2007vb,%
  Dumitru:2008wn,Fukushima:2008ya} as well as on the canonical
formalism~\cite{Dumitru:2001ux}.  One of the aims of this paper is to
establish a link, by a concrete example, between the formalism of LSZ
reduction formulas, which is usually associated with perturbation
theory only, and non-perturbative tunneling phenomena of the Schwinger
mechanism. More concretely, we want to show that the LSZ perturbative
framework automatically includes the particles produced by the
Schwinger mechanism, provided the external field is properly resummed.
Thus, this contribution does not need to be added separately by hand.

For applying the mechanism of pair creation from a classical field to
phenomenology one needs, in addition to the vacuum decay rate or
spectrum of pairs, the whole probability distribution of multiparticle
production.  In many practical applications of the Schwinger
mechanism, there has been a confusion of terminology, with both the
formulas and the concepts of the vacuum decay rate (or persistence
probability calculated by Schwinger) and the pair production rate.
The difference between the two was recently nicely discussed in
ref.~\cite{Cohen:2008wz}, where it is interpreted as a result of
temporal correlations between the produced pairs.  In fact these two
were clearly distinguished already in a classical paper by
Nikishov~\cite{Nikishov:1970br}.  In the case of the typical QED
discussion, the pair production rate is extremely small, in which case
the probability distribution of produced pairs cannot be distinguished
from a Poisson distribution.  This seems to have been the assumption
used, without any further justification, also in many QCD
phenomenological applications (see e.g.~\cite{Biro:1984cf} where this
is very explicit).  As we shall show explicitly in the following, this
assumption is not true when the pair production is not strongly
suppressed, as can typically be the case in e.g.\ heavy-ion
collisions.  Instead, the probability distribution of the produced
pairs turns out naturally to be the appropriate (Bose-Einstein or
Fermi-Dirac) quantum one~\footnote{
Note that the ``inversion of spin statistics'' discussed in 
\cite{Muller:1977mm,PauchyHwang:2009rz} refers to a formal expression
of the vacuum decay rate as an integral over the BE or FD distribution function
and not the actual probability distribution of produced particles.}. 
 With explicit expressions for the
probabilities to produce one, two, etc.\ particles, the distinction
between the vacuum decay rate (related to the probability to produce
no pairs) and the pair production rate (the expectation value of the
number of pairs produced), becomes obvious.  The fact that there is a
quantum statistical (BE or FD) correlation has long ago been realized
by some authors as a requirement that should in principle be built
into Monte Carlo event
generators~\cite{Zajc:1986sq,Lonnblad:1995mr,Lonnblad:1997kk}.  Also,
the full computation of the vacuum decay rate should encompass all the
multiparticle production processes,--because of unitarity--, including
the quantum statistics.  To our knowledge, however, an explicit
derivation of how the BE or FD correlations arise from the Schwinger
mechanism has been lacking.  Besides, the multiparticle distribution
has scarcely drawn attention in the context of the Schwinger
mechanism, probably because of the absence of experimental access.  In
fact the spectrum of multiparticle production is quite informative and
precise data of charged hadron multiplicity fluctuations are already
available in $p$-$\bar{p}$~\cite{Alner:1987wb} and
heavy-ion~\cite{Abbott:1995as} collision experiments, where a negative
binomial distribution gives a beautiful fit.  Of course, to account
for the high-energy experimental data, a simple treatment of spatially
homogeneous (i.e.\ constant in space) background fields is inadequate,
and recently, it has been shown that an inhomogeneous configuration
forming a certain number of the glasma flux tubes leads to the
negative binomial distribution~\cite{Gelis:2009wh}.  This is beyond
the scope of our current paper.

In the following, we shall study the case of scalar QED in a
time-dependent but spatially homogeneous external gauge field.  We
shall first introduce the model and derive the probability
distribution of produced pairs using the LSZ reduction formula in
\se\ref{sec:lsz}.  Then, in \se\ref{sec:bogo}, we shall rederive the
same results using canonical quantization and interpret the
calculation of \se\ref{sec:lsz} in terms of a Bogoliubov
transformation.  The discussion in \se\ref{sec:bogo} to a large degree
follows that of Tanji~\cite{Tanji:2008ku}, but takes the additional
mathematically simple step of actually writing down the whole
probability distribution (see also \cite{Kim:2008yt}).
Then, in \se\ref{sec:exact} we shall
demonstrate how this procedure works in practice using an exactly
solvable example (see also \cite{Kim:2007pm})
of a time dependent external potential, from which we
can take both the constant field and short pulse limits.  We note
that, in all our discussions, we will solve the problem for a given
external field without taking into account the interplay between the
field and the produced particles which screen the external field,
leading to plasma oscillation behavior in
time~\cite{Tanji:2008ku,Kluger:1991ib,Hebenstreit:2008ae}.

\section{LSZ derivation}
\label{sec:lsz}
We here calculate the Schwinger mechanism in terms of the LSZ
reduction formulas.  To this aim we develop a slightly modified
version of the Schwinger-Keldysh formalism to compute the generating
functional of the particle and antiparticle spectra.

\subsection{Model}
To avoid encumbering the discussion with unessential details, we
consider scalar QED, i.e.\ a charged scalar field $\phi$ coupled to an
external vector potential $A^\mu$.  Moreover, in order to simplify
things even further, we neglect any kind of self-interactions among
the scalar fields, and the coupling to the external electromagnetic
field enters only via the covariant derivatives,
$D^\mu=\partial^\mu-\rmi eA^\mu$.  Thus, the Lagrangian of this model
is:
\begin{equation}
\mathcal{L}=\left(D_\mu\phi\right)\left(D^\mu\phi\right)^*
-m^2\phi\phi^*\; .
\label{eq:lagrangian}
\end{equation}
In most of the considerations of this paper, we need not specify the
precise form of the background potential $A^\mu$.  Only in the final
section, we work out the case of an explicit example of background
electric field that leads to exact analytical results.

\subsection{Reduction formulas}
We assume that the initial state of the system does not contain any
particles or antiparticles.  However, because of the background field,
transitions to populated states are possible.  Let us consider the
following transition amplitudes,
\begin{equation}
\mathcal{M}_{m,n}(\{\p_i\},\{\q_i\})
\equiv
\big<\underbrace{\p_1\cdots\p_m}_{\mbox{particles}}\;\;
\underbrace{\q_1\cdots\q_n}_{\mbox{antiparticles}}{}_{\rm out}
\big|0_{\rm in}\big>\; ,
\end{equation}
from the vacuum to a populated state.  The conservation of electrical
charge implies that an equal number of particles and antiparticles
must be produced, i.e.\ that this general amplitude is proportional to
$\delta_{mn}$.  This transition amplitude can be obtained from the
expectation value of time-ordered products of fields:
\begin{align}
\mathcal{M}_{m,n}(\{\p_i\},\{\q_i\}) = & \int
\prod_{i=1}^m \rmd^4x_i \; \rme^{\rmi p_i\cdot x_i}(\square_{x_i}\!+m^2)
\prod_{j=1}^n \rmd^4y_j \; \rme^{\rmi q_j\cdot y_j}(\square_{y_j}\!+m^2)
\notag\\
&\times
\big<0_{\rm out}\big| {\rm T}\,
\phi(x_1)\cdots\phi(x_m)
\phi^*(y_1)\cdots\phi^*(y_n)\big|0_{\rm in}\big>\; .
\label{eq:LSZ}
\end{align}
Here the on-shell boundary condition is the vacuum one, i.e.  $p_i^0
\to \sqrt{\p_i^2+m^2}$ and $q_i^0\to \sqrt{q_i^2+m^2}$ for particles
and antiparticles, respectively.  This is adequate only if one chooses
a gauge\footnote{It is always possible to find such a gauge if the
  electrical field vanishes when time goes to infinity, a necessary
  condition to be able to unambiguously define what we mean by
  ``measuring a particle''. If one insists on using a gauge in which
  $A_\mu$ is not zero when $x^0\to +\infty$, one must replace the ordinary
  derivatives by covariant derivatives and the plane waves by gauge
  transformed plane waves in eq.~(\ref{eq:LSZ}). The mass-shell
  condition for $p_i^0$ and $q_i^0$ should also be altered by the
  non-zero background gauge field.} in which the background field
$A_\mu$ vanishes when time goes to $+\infty$.  Because the conjugate
$\phi^*(y_i)$ already takes care of antiparticle nature, $q_i^0$
should also be chosen to be positive.  Note that, in principle, each
field in this formula should be accompanied by a wave-function
renormalization factor, $Z^{-1/2}$.  However, since we do not include
any self-interactions among the fields, these factors are equal to
unity here and we can safely ignore them.

\subsection{Generating functional: definition}
All the physical quantities related to particle production in this
model can be constructed from the squared amplitudes
$\left|\mathcal{M}_{m,n}\right|^2$.  A very useful object that
contains all this information in a compact form is the generating
functional defined by~\cite{Gelis:2006yv,Gelis:2006cr}
\begin{equation}
\mathcal{F}[z,\bar{z}]
\equiv
\sum_{m,n=0}^\infty
\frac{1}{m!n!}
\int
\prod_{i=1}^m \rmd^3\p_i\;z(\p_i)
\prod_{j=1}^n \rmd^3\q_j\;\bar{z}(\q_j)\;
\Big|\mathcal{M}_{m,n}(\{\p_i\},\{\q_i\})\Big|^2\; .
\label{eq:def_gen}
\end{equation}
In this functional, $z$ and $\bar{z}$ are two functions defined over
the 1-particle momentum space (unlike what the notation may suggest,
they are independent and not complex conjugates of each other).

If one sets the functions $z$ and $\bar{z}$ to constants equal to
unity, one gets,
\begin{equation}
{\cal F}[1,1]=\sum_{m,n=0}^\infty P_{m,n}\; ,
\label{eq:prob}
\end{equation}
where $P_{m,n}$ is the total probability to have $m$ particles and $n$
antiparticles in the final state.  From unitarity, the sum of all
these probabilities must be equal to one, hence
\begin{equation}
{\cal F}[1,1]=1\; .
\label{eq:unitarity}
\end{equation}
This is an important constraint on the generating functional
$\mathcal{F}[z,\bar{z}]$, that leads to significant simplification in
the computation of inclusive observables.

Assuming that this generating functional is known, one can obtain the
single inclusive particle spectrum as
\begin{equation}
\frac{\rmd N_1^+}{\rmd^3\p}
=
\left.\frac{\delta\mathcal{F}[z,\bar{z}]}{\delta z(\p)}
\right|_{z=\bar{z}=1}\; ,
\end{equation}
the single inclusive antiparticle spectrum as
\begin{equation}
\frac{\rmd N_1^-}{\rmd^3\q}
=
\left.\frac{\delta\mathcal{F}[z,\bar{z}]}{\delta \bar{z}(\q)}
\right|_{z=\bar{z}=1}\; ,
\end{equation}
and the double inclusive particle-particle spectrum as
\begin{equation}
\frac{\rmd N_2^{++}}{\rmd^3\p_1 \rmd^3\p_2}
=
\left.\frac{\delta^2\mathcal{F}[z,\bar{z}]}
{\delta z(\p_1) \, \delta z(\p_2)}\right|_{z=\bar{z}=1}\; .
\end{equation}
Other combinations of inclusive 2-particle spectra are given by
\begin{equation}
\frac{\rmd N_2^{--}}{\rmd^3\q_1 \rmd^3\q_2}
=
\left.\frac{\delta^2\mathcal{F}[z,\bar{z}]}
{\delta \bar{z}(\q_1) \, \delta \bar{z}(\q_2)}
\right|_{z=\bar{z}=1}\quad,\quad
\frac{\rmd N_2^{+-}}{\rmd^3\p \, \rmd^3\q}
=
\left.\frac{\delta^2\mathcal{F}[z,\bar{z}]}
{\delta z(\p) \, \delta \bar{z}(\q)}\right|_{z=\bar{z}=1}\; .
\end{equation}
Note that from their definitions, these two particle spectra are
normalized so that their integrals over $\p$ and $\q$ are,
respectively,
\begin{equation}
 \begin{split}
\int \rmd^3\p_1 \rmd^3\p_2\;
\frac{\rmd N_2^{++}}{\rmd^3\p_1 \rmd^3\p_2}
&=
\big<{\bs N}^+({\bs N}^+-1)\big>\; ,\\
\int \rmd^3\q_1 \rmd^3\q_2\;
\frac{dN_2^{--}}{\rmd^3\q_1 \rmd^3\q_2}
&=
\big<{\bs N}^-({\bs N}^--1)\big>\; ,\\
\int \rmd^3\p \, \rmd^3\q\;
\frac{\rmd N_2^{+-}}{\rmd^3\p \, \rmd^3\q}
&=
\big<{\bs N}^+{\bs N}^-\big>\; ,
 \end{split}
\label{eq:N2++int}
\end{equation}
where ${\bs N}^\pm$ denote the number operator for particles and
antiparticles in the final state respectively.  In terms of the total
probability introduced in \eq(\ref{eq:prob}) we will see that
$\big<{\bs N}^+\big>=\sum_{m,n}mP_{mn}$,
$\big<{\bs N}^-\big>=\sum_{m,n}nP_{mn}$,
$\big<{\bs N}^+({\bs N}^+-1)\big>=\sum_{m,n}m(m-1)P_{mn}$, etc, for
which one can find a justification in the
appendix~\ref{app:multiplicity}.  What these equations mean in the
$++$ and $--$ cases is that our 2-particle spectra are defined by
summing over all possible pairs of \textsl{distinct} particles in
every event.  When summed over all momenta in a given event, this
leads to $N^\pm(N^\pm-1)$ where $N^\pm$ is the multiplicity of
particles (resp.\ antiparticles) in that event.  Naturally, this
requirement of taking distinct particles has no incidence on the $+-$
case -- since charged particles are always distinct from the
corresponding antiparticles --, which explains the last of
\eqs(\ref{eq:N2++int}).

\subsection{Generating functional: computation}
Let us now proceed to the actual computation of the generating
functional $\mathcal{F}[z,\bar{z}]$.  There is usually no closed form
answer for this object.  Since we are neglecting the self-interactions
of the fields $\phi$ in our model, however, this becomes a much
simpler calculation.  It has been shown
before~\cite{Gelis:2006yv,Gelis:2006cr} that the generating functional
$\mathcal{F}[z,\bar{z}]$ is the sum of all the
vacuum-vacuum\footnote{Vacuum-vacuum graphs are diagrams that have no
  external legs with respect to $\phi$.} graphs in a slightly modified
version of the Schwinger-Keldysh formalism~\cite{Baltz:2001dp}, where
the off-diagonal components $G_{+-}^0$ and $G_{-+}^0$ of the free
propagator are altered by the functions $z$ or $\bar{z}$.  Explicitly,
the propagators read:
\begin{equation}
 \begin{split}
G_{++}^0(p) &= \frac{\rmi}{p^2-m^2+\rmi\epsilon}\; ,\qquad
G_{--}^0(p) = \frac{-\rmi}{p^2-m^2-\rmi\epsilon}\; ,\\
G_{+-}^0(p) &= 2\pi\theta(-p^0)\,z(\p)\,\delta(p^2-m^2)\; ,\\
G_{-+}^0(p) &= 2\pi\theta(+p^0)\,\bar{z}(-\p)\,\delta(p^2-m^2)\; .
 \end{split}
\label{eq:SK-props}
\end{equation}
As one can see, the off-diagonal free propagators are simply
multiplied by $z(\p)$ and $\bar{z}(\p)$ respectively.  Given these
propagators, the rules for calculating $\mathcal{F}[z,\bar{z}]$ are
straightforward:
\begin{itemize}
\item[{\bf i.}] Draw all the vacuum-vacuum diagrams at the desired
  order.  There are simply connected and multiply connected graphs.
  However, one can always limit the calculation to the simply
  connected ones, and then exponentiate the result in order to obtain
  the full result that also includes the multiply connected ones.
\item[{\bf ii.}] For a given graph, sum over all the possible ways to
  assign $+$ or $-$ signs to the vertices.
\item[{\bf iii.}] A $-$ vertex is the complex conjugate of a $+$
  vertex.  Let us denote by ${\cal V}(A)$ the value of the coupling of
  $\phi,\phi^*$ to the background field in a $+$ vertex. Note that
  this 'potential' is not simply $A_\mu$ itself, since there are both
  a $e(\partial_\mu\phi) \phi^* A^\mu$ and a $e^2 \phi \phi^* A_\mu
  A^\mu$ couplings -- however, we will not need its detailed
  expression in the following. The corresponding $-$ vertex is ${\cal
    V}^*(A)=-{\cal V}(A)$ (this identity follows from the hermiticity
  of the Lagrangian).
\item[{\bf iv.}] Connect these vertices with the propagators defined
  in \eq(\ref{eq:SK-props}). 
\end{itemize}

\begin{figure}[htbp]
\begin{center}
\includegraphics[width=3cm]{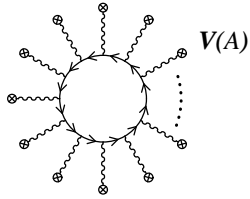}
\end{center}
\caption{\label{fig:loop}Topology of the connected vacuum-vacuum
  diagrams that contribute to $\ln\mathcal{F}$.  The solid line
  denotes the free propagator of the charged scalar field $\phi$ (the
  arrow indicates the direction of the flow of positive electric
  charge).  The wavy line terminated by a circled cross denotes the
  background gauge potential ${\cal V}(A)$.  Note that, in scalar QED,
  the background 'potential' is not simply $A_\mu$ itself, since there
  are both a $e(\partial_\mu\phi) \phi^* A^\mu$ and a $e^2 \phi \phi^*
  A_\mu A^\mu$ couplings.  The wavy line represents the sum of these
  two contributions.}
\end{figure}

\begin{figure}[htbp]
\begin{center}
\includegraphics[width=6cm]{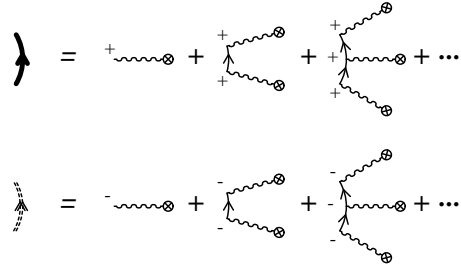}
\end{center}
\caption{\label{fig:block}Building blocks for the summation of field
  insertions having a fixed Schwinger-Keldysh vertex assignment.}
\end{figure}

\begin{figure}[htbp]
\begin{center}
\includegraphics[width=3cm]{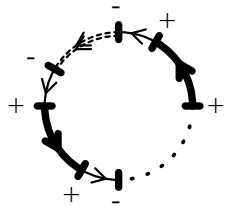}
\end{center}
\caption{\label{fig:resum}Block decomposition of the double summation
  over the number of background field insertions and the $\pm$
  assignments at the vertices.}
\end{figure}

Step {\bf i} is trivial:  there is only one topology of simply
connected vacuum-vacuum graph in our model.  These are the graphs made
of a single closed loop, embedded with the background field $A^\mu$,
as illustrated in \fig\ref{fig:loop}.  We must sum over the number of
insertions of the background potential $A_\mu$ (from zero to infinite
insertions), and for each of these insertions we must sum over the
type $+$ and $-$ for the corresponding vertex.  This double summation
can be organized in blocks, as illustrated in \figs\ref{fig:block} and
\ref{fig:resum}.


If we denote by $\mathcal{T}_+$ the sum of graphs in the first line of
\fig\ref{fig:block} and by $\mathcal{T}_-$ the sum of graphs on the
second line of the same figure, we can take the remaining steps
{\bf ii}, {\bf iii}, and {\bf iv} and the sum of the vacuum-vacuum
diagrams contributing to $\ln\mathcal{F}$ can be written as
\begin{equation}
\ln{\cal F}[z,\bar{z}]
=
\mbox{constant}
+
\sum_{n=1}^\infty
\frac{1}{n}
\;{\rm tr}\,\Big[\mathcal{T}_+ G_{+-}^0
\mathcal{T}_- G_{-+}^0\Big]^n\; ,
\label{eq:lnF1}
\end{equation}
where the trace symbol (${\rm tr}\,(\cdots)$) denotes an integration
over the space-time coordinates (not represented explicitly in the
formula) of all the vertices.  The first term in this formula, that we
simply denoted ``constant'' but did not write explicitly, is
independent of $z$ and $\bar{z}$.  It is made of all the graphs in
which all the vertices are of type $+$ or all of type $-$ (and thus
cannot contain $z$ nor $\bar{z}$ since these come with the
$G_{\pm\mp}^0$ propagators).  In the second term of this formula, the
index $n$ represents the number of the block consisting of one $+-$
and one $-+$ transitions, and the factor $1/n$ is a symmetry factor
since we can rotate the graph \emph{by one block} without altering it.
Note that the index $n$ gives the order in $z$ and $\bar{z}$ of the
corresponding term\footnote{We see explicitly here that the order in
  $z$ of a given term is the same as its order in $\bar{z}$, which
  reflects the fact that particles and antiparticles can be created
  only in pairs.}.

It is trivial to perform explicitly the summation in
\eq(\ref{eq:lnF1}) to find,
\begin{equation}
\ln\mathcal{F}[z,\bar{z}]
=
\mbox{constant}
-
{\rm tr}\,\ln\Big[1-\mathcal{T}_+ G_{+-}^0
\mathcal{T}_- G_{-+}^0\Big]\; .
\label{eq:lnF2}
\end{equation}
The constant term that we did not write explicitly can be determined
without any calculation so that $\ln\mathcal{F}[1,1]=0$, as required
from unitarity~\nr{eq:unitarity}.  Therefore, we have
\begin{equation}
\mathcal{F}[z,\bar{z}]
=
\frac
{\exp\left(-{\rm tr}\,\ln\Big[1-\mathcal{T}_+ G_{+-}^0
 \mathcal{T}_- G_{-+}^0\Big]\right)}
{\exp\left(-{\rm tr}\,\ln\Big[1-\mathcal{T}_+ G_{+-}^0
 \mathcal{T}_- G_{-+}^0\Big]_{z=\bar{z}=1}\right)}\; .
\label{eq:F-final}
\end{equation}
Although fairly formal, this formula contains all we need to know
about the particle production by an external electromagnetic field in
scalar QED.

In order to simplify the subsequent discussion, let us restrict
ourselves to background electric fields that do not depend on the
position $\x$.  In this case it is always possible to choose a gauge
in which the background vector potential $A_\mu(x)$ is also
independent of $\x$, and its Fourier transform is proportional to a
delta function $\delta(\k)$ as far as its dependence on the spatial
components of the momentum is concerned.  In this particular case, the
2-point function $\mathcal{T}_+ G_{+-}^0 \mathcal{T}_- G_{-+}^0$ has
the same entering and outgoing momenta.  In order to make more
explicit the $z$ and $\bar{z}$ dependence, let us introduce the
notation:
\begin{equation}
\Big[\mathcal{T}_+ G_{+-}^0 \mathcal{T}_- G_{-+}^0\Big]_\p
\equiv
z(\p) \bar{z}(-\p) L_\p\; .
\label{eq:L}
\end{equation}
The important points here are that $L$ does not contain $z$ and
$\bar{z}$, and that the functions $z$ and $\bar{z}$ carry the same
momentum up to a relative sign\footnote{Physically, the building block
  in \eq(\ref{eq:L}) is the amplitude squared for producing a single
  particle-antiparticle pair.  Since the background field is uniform
  in space, the total momentum of this pair must be zero.  Hence the
  opposite sign for the momentum argument of $z$ and $\bar{z}$.}.
With this compact notation:
\begin{equation}
\mathcal{F}[z,\bar{z}]
=
\frac
{\exp\left(-{\rm tr}\,\ln\Big[1-z\bar{z}L\Big]\right)}
{\exp\left(-{\rm tr}\,\ln\Big[1-L\Big]\right)}\; .
\label{eq:F-final1}
\end{equation}
From now on, it is simpler to calculate the trace in momentum space.
Indeed, when the background potential is space independent, a unique
momentum $\p$ runs around the loop.

To close this subsection, let us mention a generic property of the
trace that appears in \eq(\ref{eq:F-final1}).  Strictly speaking, when
the background electric field is independent of the location in space,
this trace exhibits a factor $(2\pi)^3\delta({\bs 0})$ in momentum
space.  This factor should be interpreted as the volume $V$ of the
system\footnote{In order to check this, one can make the background
  electric field slightly space dependent, so that it has a compact
  support in space.  One sees now that all the integrals are finite,
  and that the single particle spectrum is proportional to the size of
  the region where the background field is non-zero.}, and its
presence is an indication that the particle spectra are proportional
to the overall volume.

\subsection{Relation with wave propagation in the field $A_\mu$}
So far, we have not attempted to calculate the object $L$ that appears
in the generating functional.  Since it is built from the
$\mathcal{T}_\pm$, which are Feynman (time-ordered) propagators
amputated of their external legs, it is clear that $L$ is related to
the propagation of small fluctuations over the background electric
field.  However, knowing that $L$ is related to the Feynman propagator
is inconvenient for practical calculations because this propagator
obeys complicated boundary conditions.  In practice, one should try to
rewrite $L$ in terms of propagators that obey simpler boundary
conditions, like the retarded propagator.

Let us start from the equation that defines $\mathcal{T}_+$ to rewrite
it into the retarded quantities.  The resummation that leads to
$\mathcal{T}_+$ can be summarized by the following Lippmann-Schwinger
equation:
\begin{equation}
\mathcal{T}_+=\mathcal{V}+\mathcal{V} G_{++}^0 \mathcal{T}_+\; ,
\end{equation}
where $\mathcal{V}$ is the sum of the two couplings to the background
field (the derivative coupling to a single $A_\mu$ and the
non-derivative coupling to $A_\mu A^\mu$).  We do not need to specify
more what $\mathcal{V}$ is.  Note that we could have written the
equation in a slightly different form:
\begin{equation}
\mathcal{T}_+=\mathcal{V}+\mathcal{T}_+ G_{++}^0 \mathcal{V}\; .
\end{equation}
(This just amounts to starting the expansion from the other end-point
of the propagator.)  Concerning $\mathcal{T}_-$, it is sufficient to
note that it is the complex conjugate of $\mathcal{T}_+$.

The amputated retarded propagator $\mathcal{T}_{_R}$ is defined from
the same equation, but the free Feynman propagator $G_{++}^0$ is
replaced by the free retarded propagator:
\begin{equation}
\mathcal{T}_{_R}
=
\mathcal{V}+\mathcal{V} G_{_R}^0 \mathcal{T}_{_R}
=
\mathcal{V}+\mathcal{T}_{_R} G_{_R}^0 \mathcal{V}\; ,
\label{eq:LS-R}
\end{equation}
where the free retarded propagator $G_{_R}^0(p)$ is defined as
\begin{equation}
G_{_R}^0(p)=\frac{\rmi}{p^2-m^2+\rmi p^0\epsilon}\; .
\end{equation}
In order to express $\mathcal{T}_+$ in terms of $\mathcal{T}_{_R}$,
the first step is to relate the Feynman and the retarded propagators.
This is done via the following well-known relationship:
\begin{equation}
G_{++}^0 = G_{_R}^0 + \rho_-\; ,
\end{equation}
where $\rho_-$ is a 2-point function whose definition in momentum
space is
\begin{equation}
\rho_\pm(p) \equiv 2\pi \theta(\pm p^0)\,\delta(p^2-m^2)\; .
\end{equation}
(Note that $\rho_-$ is nothing but $G_{+-}^0$ with $z=1$.)
From the above equations, we arrive trivially at
\begin{equation}
(1-\mathcal{V} G_{_R}^0 -\mathcal{V}\rho_-)\mathcal{T}_+
=
(1-\mathcal{V}G_{_R}^0)\mathcal{T}_{_R}\; ,
\end{equation}
and subsequently at
\begin{equation}
(1-\underbrace{(1-\mathcal{V}G_{_R}^0)^{-1}\mathcal{V}}_{\mathcal{T}_{_R}}
\,\rho_-)\mathcal{T}_+
=
\mathcal{T}_{_R}\; .
\end{equation}
Thus, we have
\begin{equation}
\mathcal{T}_+
=
(1-\mathcal{T}_{_R}\rho_-)^{-1}\,\mathcal{T}_{_R}\; .
\label{eq:t_pl}
\end{equation}
Similarly, one can prove,
\begin{equation}
\mathcal{T}_+
=
\mathcal{T}_{_R}\,(1-\rho_- \mathcal{T}_{_R})^{-1}\; .
\label{eq:t_pl2}
\end{equation}
In fact it is easy to confirm that \eqs(\ref{eq:t_pl}) and
(\ref{eq:t_pl2}) are equivalent by expanding the inverse quantity in
terms of $\mathcal{T}_{_R}\rho_-$ in \eq(\ref{eq:t_pl}) and
$\rho_-\mathcal{T}_{_R}$ in \eq(\ref{eq:t_pl2}).  Taking the complex
conjugate of \eq(\ref{eq:t_pl}), we get,
\begin{equation}
\mathcal{T}_-=\mathcal{T}_+^*=
(1-\mathcal{T}_{_R}^*\rho_-)^{-1} \mathcal{T}_{_R}^*\; .
\end{equation}
($\rho_-$ is purely real.)  Multiplying \eq(\ref{eq:t_pl2}) by
$\rho_-$ on the right, we finally obtain,
\begin{equation}
\mathcal{T}_+\rho_-
=
\mathcal{T}_{_R}\,(1-\rho_- \mathcal{T}_{_R})^{-1}\,\rho_-
=
\mathcal{T}_{_R}\,\rho_-\,(1-\mathcal{T}_{_R}\rho_-)^{-1}\; .
\end{equation}
Combining everything, we obtain the following expression for
$z\bar{z}L$:
\begin{equation}
z\bar{z}L
=
\mathcal{T}_+ z\rho_- \mathcal{T}_- \bar{z}\rho_+
=
\mathcal{T}_{_R} z\rho_- (1-\mathcal{T}_{_R}\rho_-)^{-1}
(1-\mathcal{T}_{_R}^* \rho_-)^{-1}\mathcal{T}_{_R}^* \bar{z}\rho_+\; .
\label{eq:L-retarded}
\end{equation}
Thus we have managed to replace all the Feynman propagators in $L$ by
retarded ones. The price to pay for this transformation is that we
have now an expression that is no longer bilinear in the propagators,
but has terms at any order $\ge 2$. As we shall see now, this apparent
complication actually disappears thanks to an identity reminiscent of
the optical theorem.

The Lippmann-Schwinger equation for $\mathcal{T}_{_R}$ is given in
\eq(\ref{eq:LS-R}). For $\mathcal{T}_{_R}^*$, it reads:
\begin{equation}
\mathcal{T}_{_R}^* = -\mathcal{V} - \mathcal{V} G_{_R}^{0*}
\mathcal{T}_{_R}^*
=
-\mathcal{V} - \mathcal{T}_{_R}^* G_{_R}^{0*}\mathcal{V}\; .
\end{equation}
Here, we have used the fact that, in a unitary theory (like scalar QED
with a real background potential), we have
$\mathcal{V}^*=-\mathcal{V}$.  Adding up the equations for
$\mathcal{T}_{_R}$ and $\mathcal{T}_{_R}^*$, we first obtain
\begin{equation}
\mathcal{T}_{_R}+\mathcal{T}_{_R}^*
=-\mathcal{T}_{_R}^* G_{_R}^{0*} \mathcal{V}
+\mathcal{V} G_{_R}^0 \mathcal{T}_{_R} \; .
\end{equation}
Finally, we can eliminate $\mathcal{V}$ in the right hand side of this
equation, in favor of $\mathcal{T}_{_R}$ or $\mathcal{T}_{_R}^*$. This
leads easily to:
\begin{equation}
\mathcal{T}_{_R}+\mathcal{T}_{_R}^*
=
-\mathcal{T}_{_R}^*\Big[G_{_R}^0+G_{_R}^{0*}\Big]{\cal T}_{_R}
=
-\mathcal{T}_{_R}^*\Big[\rho_+-\rho_-\Big]\mathcal{T}_{_R} \; .
\label{eq:optical}
\end{equation}
Note that this relation is a variant of the optical
theorem\footnote{This is why the relation
  $\mathcal{V}^*=-\mathcal{V}$, that is the manifestation of unitarity
  in this calculation, is crucial in order to obtain
  \eq(\ref{eq:optical}).}  applied to a 2-point function.  The left
hand side is equal to the discontinuity of the 2-point function across
the real energy axis, and the right hand side gives the expression of
this discontinuity in terms of cut graphs.  Thanks to this
relationship, it is now straightforward to check that
\begin{equation}
(1-\mathcal{T}_{_R}^*\rho_-)(1-\mathcal{T}_{_R}\rho_-)
=
1+\mathcal{T}_{_R}^* \rho_+ \mathcal{T}_{_R} \rho_-\; .
\label{eq:optical-1}
\end{equation}

By combining \eqs(\ref{eq:L-retarded}) and (\ref{eq:optical-1}), we
easily arrive at the following simplification;
\begin{equation}
 \begin{split}
 1-z\bar{z}L &= 1-\mathcal{T}_{_R} z\rho_- (1+\mathcal{T}_{_R}^*\rho_+
  \mathcal{T}_{_R}\rho_-)^{-1} \mathcal{T}_{_R}^* \bar{z}\rho_+ \\
 &= (1+\mathcal{T}_{_R}^*\rho_+\mathcal{T}_{_R}\rho_-)^{-1}
  \Bigl[ 1 - (z\bar{z}-1)\mathcal{T}_{_R}\rho_-
  \mathcal{T}_{_R}^*\rho_+ \Bigr] \;,
 \end{split}
\end{equation}
which leads to the generating functional:
\begin{equation}
\mathcal{F}[z,\bar{z}]
=
\exp\Big(
 -{\rm tr}\,\ln\Big[1-(z\bar{z}-1)\mathcal{T}_{_R}
 \rho_- \mathcal{T}_{_R}^* \rho_+
\Big]\Big)\; .
\end{equation}
Here again, thanks to the fact that the background field is uniform,
all the factors inside the logarithm share a single spatial momentum
$\p$. The function $z$ has argument $\p$ and the $\bar{z}$ is
evaluated at $-\p$.

At this point, we see that all the properties of the distribution of
produced particles are determined by a single quantity, namely the
amputated retarded propagator $\mathcal{T}_{_R}$ of a scalar particle
on top of the background field $A^\mu$.  Before going further, it may
be useful to rewrite the combination
$z\bar{z}\,\mathcal{T}_{_R}\rho_-\mathcal{T}_{_R}^* \rho_+$ with all
the momentum dependence:
\begin{equation}
  \Big[z\bar{z}\,\mathcal{T}_{_R}\rho_- \mathcal{T}_{_R}^*
  \rho_+\Big]_{p,q}
  =\bar{z}(-\q)\rho_+(q)
  \int \frac{\rmd^4k}{(2\pi)^4}\;z(\k)\;
  \mathcal{T}_{_R}(p,k)\rho_-(k)
  \mathcal{T}_{_R}^*(k,q)\; .
\label{eq:final-R}
\end{equation}
This formula is completely general. Note that $\mathcal{T}_{_R}^*$ is
the same quantity as $\mathcal{T}_{_R}$, in which all the retarded
propagators are replaced by advanced ones and ${\cal V}(A)$ is
replaced by ${\cal V}^*(A)$. 

In eq.~(\ref{eq:final-R}), the momenta $k$ and $q$ are forced to be on
the in-vacuum mass-shell, since they appear inside the distribution
$\rho_\pm$. However, in the case of $k$, it turns out to be simpler to
have a momentum variable (which we denote here by $\tilde{k}$)
that obeys the mass-shell condition imposed
by the non-zero gauge potential at $x^0\to -\infty$.  Let us denote
$G_{_R}^{\infty}$ the retarded propagator evaluated in the presence
of the (constant) background field
\begin{equation}
A^{\infty}_\mu\equiv\lim_{x_0\to -\infty}A_\mu(x)\; .
\end{equation}
Since there is no electrical field at $x^0\to -\infty$, the gauge
field is a pure gauge in this limit:
\begin{equation}
A_\mu^{\infty}
=
\partial_\mu \chi(x)\; ,
\end{equation}
and the propagator $G_{_R}^{\infty}$ is simply obtained by a gauge
transformation from the vacuum propagator $G_{_R}^0$:
\begin{equation}
G_{_R}^{\infty}(x,y)
=e^{ie\chi(x)}\,G_{_R}^0(x,y)\,e^{-ie\chi(y)}\; .
\end{equation}
Let us now rewrite the combination ${\cal T}_{_R}\rho_- \mathcal{T}_{_R}^*$ that
appears in  eq.~(\ref{eq:final-R}) 
in terms of the corresponding expressions ${\cal T}_{_R}^\infty$
and $\rho_-^\infty$ which are naturally functions of the modified
mass shell momentum $\tilde{k}$. 
This can be done by writing eq.~(\ref{eq:final-R})  as
\begin{equation}
{\cal T}_{_R}\rho_- {\cal T}_{_R}^*
=
\underbrace{{\cal T}_{_R}G_{_R}^0(G_{_R}^{\infty})^{-1}}_{{\cal T}_{_R}^\infty}
\underbrace{G_{_R}^{\infty}(G_{_R}^0)^{-1}\rho_-\left((G_{_R}^0)^{-1}G_{_R}^{\infty}\right)^*}_{\rho_-^\infty}
\underbrace{\left((G_{_R}^\infty)^{-1}G_{_R}^0 {\cal T}_{_R}\right)^*}_{({\cal T}_{_R}^\infty)^*}\; .
\end{equation}
Note that because 
\begin{equation}
(D_x^\infty + m^2)G_{_R}^{\infty}(x,y) = \delta(x-y) \;,
\end{equation}
where $D_x^\infty$ is the covariant derivative
constructed with the asymptotic field $A_\mu^\infty$ at $y^0\to -\infty$,
and $\rho_-^\infty(x,y)$ is still translationally invariant, it now projects
momenta to the mass shell in presence of the background field at $x^0\to -\infty$.
Since this gauge field is a pure gauge, it is easy to write the corresponding
mass-shell conditions imposed by $\rho_+^\infty(\tilde{k})$ and 
$\rho_-^\infty(\tilde{k})$:
\begin{equation}
(\tilde{k} \pm eA^\infty)^2=m^2\; ,
\label{eq:MS-A}
\end{equation}
where the signs $\pm$ are to be chosen for the
positive and negative energy solutions respectively.
In a gauge where $A^0=0$, as we shall chose later on, these read:
\begin{equation}
\tilde{k}^0= E_{\k}^{\rm in}\; \quad\mbox{and\ \ }\;\tilde{k}^0= -E_{-\k}^{\rm in}
  ,\quad\mbox{where\ \ }E_{\k}^{\rm in}\equiv \sqrt{(\k+e{\bs A}^\infty)^2+m^2}\; .
\end{equation}

Therefore, eq.~(\ref{eq:final-R}) can be rewritten as
\begin{eqnarray}
  \Big[z\bar{z}\,\mathcal{T}_{_R}\rho_- \mathcal{T}_{_R}^*
  \rho_+\Big]_{p,q}
  &=&
  \bar{z}(-\q)\rho_+(q)
  \int \frac{\rm d^4 \tilde{k}}{(2\pi)^4}\;z(\k)\;
  \mathcal{T}_{_R}^\infty(p,\tilde{k})\rho^\infty_-(\tilde{k})
  (\mathcal{T}_{_R}^\infty(\tilde{k},q))^*
  \nonumber\\
  &=&
  z(\p)\bar{z}(-\q)\rho_+(q)
  \int \frac{\rmd^4 \tilde{k}}{(2\pi)^4}\,
  \left|\mathcal{T}_{_R}^\infty(p,-\tilde{k})\right|^2\;\rho_+^\infty(\tilde{k})
\; .
\label{eq:final-R-1}
\end{eqnarray}
In the second line, we have changed $\tilde{k}\to -\tilde{k}$, and we
have exploited the fact that $\mathcal{T}_{_R}^\infty(p,-\tilde{k})$
is proportional to $\delta(\p+\k)$ in a uniform background field.
Note that thanks to the constraints provided by the $\rho_+(q)$ and
$\rho_+^\infty(\tilde{k})$ factors, the energies $\tilde{k}^0$ and
$q^0$ are both positive. However, they obey different mass-shell
conditions. The outgoing particle\footnote{ Recall that we will need
  the trace of $ \Big[z\bar{z}\,\mathcal{T}_{_R}\rho_-
  \mathcal{T}_{_R}^* \rho_+\Big]$ and thus $q$ will be equal to the
  momentum of the produced particle $p$.}  energy $q^0$ follows the
in-vacuum dispersion relation, while $\tilde{k}^0$ obeys the
dispersion relation in the presence of the background field
$A_\mu^\infty$.

For practical calculations of $\mathcal{T}_{_R}^\infty(p,-\tilde{k})$, it is
best to relate this quantity to the Fourier coefficients of a plane
wave propagating on top of the background field. Since
$\mathcal{T}_{_R}^\infty$ is obtained by amputating the retarded
propagator $G_{_R}$ with $(G_{_R}^0)^{-1}$ on the right and with
$(G_{_R}^\infty)^{-1}$ on the left, we can immediately write:
\begin{eqnarray}
\mathcal{T}_{_R}^\infty(p,-\tilde{k})
&=&
\int d^4x\; e^{\rmi p\cdot x}\,(\square_x+m^2)
\underbrace{\int d^4y\; e^{\rmi \tilde{k}\cdot y}\,(D^{\infty2}_y+m^2)\;
G_{_R}(x,y)}_{\eta_\k(x)}
\nonumber\\
&=&\lim_{x^0\to+\infty}\int
  \rmd^3\x\; \rme^{\rmi p\cdot x}\;
 (\partial_{x_0}-\rmi E_\p^{\textrm{out}})\;\eta_{\k}(x)\; .
\label{eq:T_R}
\end{eqnarray}
In these formulas $E_\p^{\rm out}$ is the vacuum on-shell energy
$E_\p^{\rm out}\equiv \sqrt{\p^2+m^2}$. Since the propagator
$G_{_R}(x,y)$ is a Green's function of the operator $D^2_x+m^2$ (now,
the covariant derivative $D^\mu$ is defined with the full background
field, not just its asymptotic value in the past),
\begin{equation}
\big[D^2_x+m^2\big]G_{_R}(x,y)=\delta^4(x-y)\; ,
\end{equation}
it is easy to check that $\eta_{\k}(x)$ obeys the following equation
of motion:
\begin{equation}
\big[D^2_x+m^2\big]\eta_{\k}(x) = 0\; ,
\label{eq:EOM-BC}
\end{equation}
provided that $\tilde{k}$ obeys the negative energy mass-shell condition 
(\ref{eq:MS-A}).  In
order to find the boundary condition when $x^0\to -\infty$ for
$\eta_\k(x)$, we can replace the full propagator $G_{_R}(x,y)$ in
eq.~(\ref{eq:T_R}) by the propagator $G_{_R}^\infty(x,y)$ that resums
only the asymptotic field $A^\infty$,
\begin{eqnarray}
\eta_\k(x)&\empile{=}\over{x^0\to-\infty}&
\int d^4y\; e^{\rmi \tilde{k}\cdot y}\,(D^{\infty2}_y+m^2)\;
G_{_R}^\infty(x,y)
\nonumber\\
&=& e^{\rmi \tilde{k} \cdot x}\; .
\label{eq:etaic}
\end{eqnarray}
In order to obtain the final formula, we have used:
\begin{eqnarray}
(D^{\infty2}_y+m^2)\;G_{_R}^\infty(x,y)
=
\delta(x-y)\; .
\end{eqnarray}
Thus, we see that the initial condition for $\eta_\k(x)$ is a plane
wave, with a momentum $\tilde{k}$ that obeys the mass-shell condition of
eq.~(\ref{eq:MS-A}).

In a uniform background field, we can simplify a bit the notations by
writing
\begin{equation}
\mathcal{T}_{_R}^\infty(p,-k)
\equiv -2\rmi E_\p^{\textrm{out}} (2\pi)^3 \delta(\p+\k)\;
 \beta_\p\; ,
\end{equation}
so that
\begin{equation}
\Big[z\bar{z}\,\mathcal{T}_{_R}^\infty\rho_-^\infty
(\mathcal{T}_{_R}^\infty)^* \rho_+\Big]_{p,q}
=
2E_\p^{\textrm{out}}\, (2\pi)^3\delta(\p-\q)\,z(\p)\bar{z}(-\q)\,
\rho_+(q)\,  \left|\beta_\p\right|^2 \; .
\label{eq:beta}
\end{equation}
The only quantity
that we need to determine in order to fully solve the problem is the
coefficient $\beta_\p$.  This is obtained by solving the equation of
motion (\ref{eq:EOM-BC}), with a plane wave initial condition when
$x^0\to-\infty$.  We note that the initial plane wave is chosen as an
antiparticle-like one in \eq(\ref{eq:etaic}) and projected into a
particle-like one in \eq(\ref{eq:T_R}), the intuitive meaning of which
will be clear in discussions in \se\ref{sec:bogo}.

\subsection{Multiparticle spectra}
Let us now use \eqs(\ref{eq:final-R}) and (\ref{eq:beta}) in order to
obtain results about the spectra of the produced particles.  From now
on, let us simply denote $E_\p^{\textrm{out}}$ as $E_\p$ in this
section, for we have chosen the definition of $\beta_\p$ so that
$E_\p^{\textrm{in}}$ will never appear in the expressions.  We shall
wait for the next section where the difference between
$E_\p^{\textrm{in,out}}$ and the physical interpretation of $\beta_\p$
will be more explicit.  The single inclusive particle spectrum is
obtained as the first derivative of the generating functional with
respect to $z(\p)$.  We obtain
\begin{align}
  \frac{\rmd N_1^+}{\rmd^3\p}
  &=
  \frac{\delta}{\delta z(\p)}\,
  {\rm tr}\,\left[z\bar{z}\rho_+\,\mathcal{T}_{_R}^\infty\rho_-^\infty
  (\mathcal{T}_{_R}^\infty)^* \right]
  \biggr|_{z,\bar{z}=1}
  \notag\\
  &=
  \frac{V}{(2\pi)^3}\,
  \left|\beta_\p\right|^2 \; .
\label{eq:1+}
\end{align}
It should be mentioned that, in accord with the definition
(\ref{eq:def_gen}), the functional differentiations with respect to
$z(\p)$ and $\bar{z}(\q)$ are not accompanied by $(2\pi)^3$.  In the
second, we have made the trace explicit.  The final $\delta(\p-\k)$
comes from the differentiation with respect to $z(\p)$.  In the last
line, we have performed the $\rmd^4k$ integration explicitly, and we
have interpreted the (infinite) factor $(2\pi)^3 \delta({\bs 0})$ as
the volume $V$ of the system.  Naturally, the spectrum of
antiparticles is identical.  For later reference, it will be useful to
introduce more compact notations as follows:
\begin{equation}
 \begin{split}
  n_\p &\equiv \frac{\rmd N_1^+}{\rmd^3\p}
  = \frac{V}{(2\pi)^3}\,
  \left|\beta_\p\right|^2\; ,\\
  f_\p &\equiv \frac{(2\pi)^3}{V}\,{n_\p} =
  \left|\beta_\p\right|^2 \; .
 \end{split}
\label{eq:f_p}
\end{equation}
Note that $f_\p$ has the interpretation of the occupation number for
the produced particles of momentum $\p$. This is clear if we integrate 
eq.~(\ref{eq:f_p}) over the momentum and write it as
\begin{equation}
\big<{\bs N}^+ \big> = \int \frac{\rmd^3\p \, \rmd^3\x }{(2 \pi)^3}
f_\p, \label{eq:phasespace}
\end{equation} 
where the properly normalized phase space measure $\rmd^3\p \, \rmd^3\x /(2 \pi)^3$,
or restoring the Planck constant $\rmd^3\p \, \rmd^3\x /h^3$,  is explicit.

Let us now turn to the 2-particle spectra.  For two particles, we
obtain
\begin{align}
& \frac{\rmd N_2^{++}}{\rmd^3\p_1 \rmd^3\p_2}
-
\frac{\rmd N_1^+}{\rmd^3\p_1}\;\frac{\rmd N_1^+}{\rmd^3\p_2}
\notag\\
&\quad =
\frac{\delta^2}{\delta z(\p_1)\,\delta z(\p_2)}\,
{\rm tr}\,\left[
z\bar{z}\rho_+\, \mathcal{T}_{_R}^\infty\rho_-^\infty (\mathcal{T}_{_R}^\infty)^* 
z\bar{z}\rho_+\, \mathcal{T}_{_R}^\infty\rho_-^\infty (\mathcal{T}_{_R}^\infty)^* 
\right]
\biggr|_{z=\bar{z}=1}
\notag\\
&\quad = \delta(\p_1-\p_2)\, n_{\p_1} f_{\p_1} \; .
\label{eq:2++}
\end{align}
The uncorrelated part of the 2-particle spectrum shows up naturally in
this calculation, and we have absorbed it in the left hand side.  The
right hand side represents the correlated component of the 2-particle
spectrum.  As one can see, particles are correlated only if they have
identical momenta.  By integrating the previous equation over $\p_1$
and $\p_2$, and by using the first of \eqs(\ref{eq:N2++int}), we
obtain:
\begin{equation}
\big<{\bs N}^+({\bs N}^+-1)\big>-\big<{\bs N}^+\big>^2
=\int \rmd^3\p\; n_\p f_\p\; ,
\end{equation}
or equivalently
\begin{equation}
\big<{\bs N}^+{\bs N}^+\big>-\big<{\bs N}^+\big>^2
=\int \rmd^3\p\; n_\p (1+f_\p)\; 
=\int \frac{\rmd^3\p \, \rmd^3 \x }{(2\pi)^3} \; f_\p (1+f_\p)\;  .
\label{eq:variance}
\end{equation}
The form with an explicit integral over $\x$ is the one that would be applied
to a system with a phase space density that depends (slowly) on the coordinate.
The first term in the right hand side (the $1$ in $1+f_\p$) is the
answer one would obtain for a Poisson distribution.  That is, if the
probability distribution is given by
\begin{equation}
 P_{mn} = \delta_{mn}\,\frac{\rme^{-\langle{\bs N}\rangle}
  \langle{\bs N}\rangle^m}{m!} \;,
\label{eq:poisson}
\end{equation}
which defines the Poisson distribution , one would have
$\big<{\bs N}^+{\bs N}^+\big>-\big<{\bs N}^+\big>^2
=\big<{\bs N}^+\big>$ (i.e.\ the variance and the mean are
identical).  Thus, the deviations from a Poisson distribution are
contained in the term proportional to $n_\p f_\p$.
Equation~(\ref{eq:variance}) indicates that these correlations are
Bose-Einstein correlations, i.e.\ due to stimulated emission of
particles in a single quantum state.

For one particle and one antiparticle, we get:
\begin{align}
\frac{\rmd N_2^{+-}}{\rmd^3\p\, \rmd^3\q}
-\frac{\rmd N_1^+}{\rmd^3\p}\;\frac{\rmd N_1^-}{\rmd^3\q}
&=
\frac{\delta^2}{\delta z(\p)\,\delta\bar{z}(\q)}\Big[
{\rm tr}\,\left[z\bar{z}\rho_+\,\mathcal{T}_{_R}^\infty\rho_-^\infty
(\mathcal{T}_{_R}^\infty)^* \right]
\notag\\
&\quad+
{\rm tr}\,\left[
z\bar{z}\rho_+\,\mathcal{T}_{_R}^\infty\rho_-^\infty (\mathcal{T}_{_R}^\infty)^* 
z\bar{z}\rho_+\,\mathcal{T}_{_R}^\infty\rho_-^\infty (\mathcal{T}_{_R}^\infty)^*
\right]
\Big]_{z=\bar{z}=1}
\notag\\
&=
\delta(\p+\q)\;
n_\p
(1+f_\p)
\; .
\label{eq:2+-}
\end{align}
In this case, the correlation can only exist if the particle and
antiparticle have opposite spatial momenta.  Again, the integrated
form of this equation reads
\begin{equation}
\big<{\bs N}^+{\bs N}^-\big>-\big<{\bs N}^+\big>\big<{\bs N}^-\big>
=\int \rmd^3\p\; n_\p (1+f_\p)\; .
\label{eq:variance+-}
\end{equation}
One may have wondered why the particle and antiparticle have
correlations.  This can be understood by the fact that, as we noted
repeatedly, the pair of a particle and an antiparticle is created at
once so that the particle and antiparticle production preserves the
momentum conservation as well as the charge conservation.  Therefore,
the correlation~\nr{eq:2+-} reflects the particle-particle
correlation~\nr{eq:2++} with $\p_1=\p$ and $\p_2=-\q$, which explains
the delta function of spatial momenta in \eq(\ref{eq:2+-}).

\subsection{Nature of the distribution}
The results of the previous subsection suggest that the distribution
of produced particles obeys the following properties:
\begin{itemize}
\item[{\bf i.}] Two particles are correlated only if they have
  identical momenta.
\item[{\bf ii.}] A particle and an antiparticle are correlated only if
  they have opposite momenta.
\item[{\bf iii.}] In a given momentum mode, the distribution of
  produced particles follows a Bose-Einstein distribution.
\end{itemize}

Let us now present a more general justification of these results.
Because the background electric field is uniform, a unique momentum
$\p$ runs around the loop.  By using results obtained in the previous
subsections, we can rewrite the generating functional explicitly as
follows:
\begin{equation}
 \mathcal{F}[z,\bar{z}]= \exp\biggl(-V\int
  \frac{\rmd^3\k}{(2\pi)^3}
  \;\ln\Bigl[1-\bigl(z(\k)\bar{z}(-\k)-1\bigr)\,f_\k\Bigr]\biggr)\; .
\label{eq:F-2}
\end{equation}
In writing \eq(\ref{eq:F-2}), as already discussed previously in this
paper, we have assumed that the system is placed in a finite volume
$V$.  This is indeed necessary in order to have a finite particle
production rate in a constant (in space) external field.   A
consistent quantization of the system in a finite volume requires one
to specify boundary conditions\footnote{Periodic ones being the most
  convenient.}  at the edges of $V$, which leads to the momentum $\k$
being a discrete variable.  The continuum in $\k$ is recovered in the
limit $V\to \infty$.  Switching now to a notation which makes this
explicit, and remembering that  $\rmd^3\k/(2\pi)^3 = \sum_{\k}$ we can
write the generating functional as
\begin{equation}
\mathcal{F}[z,\bar{z}]
=
\prod_{\k}
\frac{1}{1+f_\k-z_\k\bar{z}_\k\,f_\k}\; ,
\end{equation}
where we use a compact notation; $z_\k\equiv z(\k)$ and
$\bar{z}_\k\equiv\bar{z}(-\k)$.  From this formula, one sees
immediately that the distributions of produced particles in the
various modes are totally uncorrelated, since the generating
functional factorizes as a product of generating functions for single
modes:
\begin{equation}
\mathcal{F}[z,\bar{z}]
=\prod_{\k}
\mathcal{F}_\k(z_\k,\bar{z}_\k)\;,\qquad
\mathcal{F}_\k(z_\k,\bar{z}_\k)\equiv
\frac{1}{1+f_\k-z_\k\bar{z}_\k\,f_\k}\; .
\end{equation}

By Taylor expanding this formula around $z_\k,\bar{z}_\k=0$, it is easy
to obtain the probability of having $m_i$ particles and $n_i$
antiparticles in the mode $i$,
\begin{equation} \label{eq:probdist}
P(\{m_i\},\{n_i\})
=
\prod_\k
\frac{\delta_{m_\k,n_\k}}{1+f_\k}
\left(\frac{f_\k}{1+f_\k}\right)^{m_\k}\; .
\end{equation}
Given the occupation numbers $f_\k$ (obtained from $\beta_\k$ by
solving the equation of motion of a plane wave over the background
field), this formula completely specifies the distribution of produced
particles and antiparticles.  Distinct modes are not correlated.  In
each mode, there must be an equal number of particles and
antiparticles.  The distribution of the particle multiplicity in the
mode $\k$ is a Bose-Einstein distribution of occupation number
$f_\k$.  A Bose-Einstein distribution is in sharp contrast to a
Poisson distribution (which would be the result in a complete absence
of correlations), since its decrease at large $m_\k$ is much slower
because of the absence of $m_\k!$ in a Poisson distribution (see the
denominator of \eq(\ref{eq:poisson})).  As a result, final states with
many particles in the same momentum mode are more likely.

\section{Bogoliubov transformation interpretation}
\label{sec:bogo}

We can interpret the results obtained in the LSZ derivation as a
Bogoliubov transformation.  To do this explicitly it is useful to
switch to canonical quantization, which we shall review here shortly
as the following manipulations are very standard ones.

From the Lagrange density of \eq(\ref{eq:lagrangian}) one obtains the
Hamiltonian of the theory (apart from the gauge part),
\begin{equation}
H = \int \rmd^3 \x \left[ \Pi \Pi^\dag + \rmi e A_0 \Pi \Phi
 - \rmi e A_0 \Pi^\dag \Phi^\dag
 + (m^2-e^2 A_0^2)\Phi \Phi^\dag + (\vec D \Phi)\cdot(\vec D \Phi)^\dag,
 \right]
\end{equation}
where $\Phi$ and $\Pi$ are operators in the Heisenberg picture
satisfying the equal-time commutation relation,
\begin{equation}
[\Phi(\x),\Pi(\y)] = \rmi \delta(\x-\y).
\end{equation}
It will be convenient for the following discussion to choose a gauge
where $A_0=0$, because in this gauge it is possible to directly
associate the time dependence of the wave function with the physical
energy of the particle\footnote{Consider for example a particle at
  rest in the vacuum: in the gauge $A_\mu=0$ its wave-function is
  $\rme^{-\rmi m x^0}$.  Performing a time-dependent gauge
  transformation with the function $M x^0/e$ will generate a
  (constant) gauge potential $A_0=-M/e$ and change the wave-function
  to $\rme^{-\rmi(m-M)x^0}$, which for $M>m$ will seemingly 
  look like a negative energy one.}.  
We shall thus work with the following Hamiltonian,
\begin{equation}
H = \int \rmd^3 \x \left[ \Pi \Pi^\dag
+ m^2 \Phi \Phi^\dag + (\vec D \Phi)\cdot(\vec D \Phi)^\dag,
 \right].
\end{equation}

The whole dynamics of the matter fields is determined by the 
equations of motion
\begin{equation} \label{eq:opeom}
 \begin{split}
\partial_0 \Phi &= \rmi\, [H,\Phi] = \Pi^\dag  \\
\partial_0 \Pi &= \rmi\, [H,\Pi] = (\vec{D}^2 - m^2)\Phi^\dag \; .
 \end{split}
\end{equation}
These can then be expressed as an equation of motion for $\Phi$ only,
but involving second order time derivatives.  The important thing to
realize is that because we are looking at a theory without
self-interactions and coupled to a classical background
field\footnote{Coupling to a quantum field would induce effective
  self-interactions through loop corrections.}, the equations of
motion of the field operators are linear;  they are in fact the same
as the classical equations of motion for the fields.  The solution to
the retarded field equations therefore contains all the information
about the relation between the field operators $\Phi$ and $\Pi$ at
$x^0\to-\infty$ and $x^0\to\infty$.  In the Heisenberg picture,
knowing the relations between the field operators is equivalent to
knowing the dynamics of the theory; in particular the whole
probability distribution of the produced particles.

We can now introduce the familiar decomposition of the field operators
in terms of creation and annihilation operators.  One can perform this
decomposition in different bases of operators; in particular the ones
that correspond to particles at $x^0\to-\infty $ (the ``in'' states)
or $x^0\to\infty $ (the ``out'' states).  Since these are just
decompositions of the same operator $\Phi$ in different bases, one
gets the equality,
\begin{equation} \label{eq:fidecomp}
 \begin{split}
\Phi(x) &= \int\frac{\rmd^3\k}{(2\pi)^3} \Biggl[
  \frac{a_{\rmin\k}}{\sqrt{2E_\k^{\textrm{in}}}} \phi^+_{\rmin\k}(x)
  +\frac{b_{\rmin\k}^\dagger}{\sqrt{2E_{-\k}^{\textrm{in}}}}\,
  \phi^-_{\rmin-\k}(x)
\Biggr] \\
&= \int\frac{\rmd^3\k}{(2\pi)^3} \Biggl[
  \frac{a_{\rmout\k}}{\sqrt{2 E_\k^{\textrm{out}}}} \phi^+_{\rmout\k}(x)
  +\frac{b_{\rmout\k}^\dagger}{\sqrt{2E_{-\k}^{\textrm{out}}}}
  \phi^-_{\rmout-\k}(x)
\Biggr]
 \end{split}
\end{equation}
Here we have denoted the dispersion relations of particles and
antiparticles in the in-state as $E_\k^{\textrm{in}}$ and that in the
out-state as $E_\k^{\textrm{out}}$.  We assume that the background
electric field is turned off adiabatically $x^0\to\pm\infty$, which
means that $A_i$ approaches a constant value $A^\textrm{in,out}_i$.
The dispersion relation for particles is then
$E_\k^{\textrm{in,out},+} = \sqrt{m^2 + (\k + e
  \mathbf{A}^\textrm{in,out})^2}$ and the one for antiparticles
$E_\k^{\textrm{in,out},-} = \sqrt{m^2 + (\k - e
  \mathbf{A}^\textrm{in,out})^2}$.  In writing \eq(\ref{eq:fidecomp})
we have used the symmetry $E_\k^- = E_{-\k}^+$ to write everything in
terms of the particle dispersion relation $E_\k^{\textrm{in,out}}
\equiv E_\k^{\textrm{in,out},+}$.  Note that the momentum label $\k$
refers to the ``canonical'' momenta, which are the variables
describing the oscillation of the wavefunction in space.  The momentum
that is actually measured in a detector is the ``kinetic'' one, which
in this case is $\k + e \mathbf{A}^\textrm{in,out}$ for particles and
$\k - e \mathbf{A}^\textrm{in,out}$ for antiparticles.  We are keeping
the notations rather general in this section.  In the physical
situation we are interested in, the only particles that are measured
are the ``out''-ones.  A convenient gauge choice, and the one adopted
in \se\ref{sec:exact} is then to take $A^\textrm{out}_i=0$, so that
one need not distinguish between the canonical and kinematical momenta
for particles in the final state--it is enough to remember that
$E_\k^{\textrm{in}}$ is different from $E_\k^{\textrm{out}}$.

The choice of basis in the decomposition \eq(\ref{eq:fidecomp}) of the
field operator is determined by the boundary conditions for the
functions $\phi^\pm_{\rmin\k}(x)$ and $\phi^\pm_{\rmout\k}(x)$.  When
we require that they approach plane waves at asymptotic times:
\begin{align}\label{eq:planewave1}
\phi^+_{\rmin\k}(x)
&= \rme^{-\rmi E_\k^\textrm{in}x^0 + \rmi \k \cdot \x}
&&\textrm{ for } x^0\to -\infty
\\ \label{eq:planewave2}
\phi^-_{\rmin\k}(x)
&= \rme^{ \rmi E_\k^\textrm{in}x^0  + \rmi \k \cdot \x}
&&\textrm{ for } x^0\to -\infty
\\\label{eq:planewave3}
\phi^+_{\rmout\k}(x)
&= \rme^{-\rmi E_\k^\textrm{out}x^0 + \rmi \k \cdot \x}
&&\textrm{ for } x^0\to +\infty
\\\label{eq:planewave4}
\phi^-_{\rmout\k}(x)
&= \rme^{ \rmi E_\k^\textrm{out} x^0 + \rmi \k \cdot \x}
&&\textrm{ for } x^0\to +\infty
\end{align}
the corresponding operators $a_\textrm{in,out}, b_\textrm{in,out}$
annihilate the in-state and out-state particles and antiparticles,
respectively.  Note that for further convenience our notation has been
chosen such that the coordinate dependence in both $\phi^+_\k(x)$ and
$\phi^-_\k(x)$ is $\rme^{ + \rmi \k \cdot \x}$ and thus the usual
negative energy plane wave $\rme^{ \rmi k \cdot x}$ corresponds to
$\phi^-_{-\k}(x)$.  The canonical commutation relation for $\Phi$ and
$\Pi$ is satisfied if\footnote{Note that in here these commutation
  relations do not include a factor $2E_\k$ as is conventional. The
  normalization used here is simpler in the case where
  $E_\k^{\textrm{in}}$ differs from $E_\k^{\textrm{out}}$.}
\begin{equation}\label{eq:abcommut}
[ a_{\rmin\k}, a^\dag_{\rmin\p} ]
=
[ b_{\rmin\k}, b^\dag_{\rmin\p} ]
=
[ a_{\rmout\k}, a^\dag_{\rmout\p} ]
=
[ b_{\rmout\k}, b^\dag_{\rmout\p} ]
= (2\pi)^3 \delta(\k-\p) \;.
\end{equation}
All the space-time dependence of the field operator $\Phi$ is in the
coefficient functions $\phi^\pm_{\textrm{in,out},\k}(x)$; the
creation and annihilation operators are time-independent.  Because the
equation of motion (\ref{eq:opeom}) for $\Phi$ is linear, the
coefficient functions $\phi^\pm_{\textrm{in,out},\k}(x)$ must each
independently satisfy the same equation.  In fact the solution to the
equation of motion $\eta_{\k}(x)$ introduced in \eq(\ref{eq:EOM-BC})
is nothing but $\phi^-_{\rmin-\k}(x)$.

The relation between the field operators at $x^0\to-\infty$ and at
$x^0\to+\infty$ is encoded in the Bogoliubov coefficients.  They are in
the transformation matrix between the in- and out-basis functions.
The solution for $\phi^-_{\rmin\k}(x)$ is again a superposition of
plane waves at $x^0 \to +\infty$.  If the background field depends only
on time, the modes of different $\k$ do not mix and  we can introduce
the Bogoliubov coefficients as the coefficients of this plane wave
decomposition by writing
\begin{equation}\label{eq:phiinplus}
\lim_{x^0\to+\infty}\phi^+_{\rmin\k}(x) = 
\sqrt{ \frac{E_\k^{\textrm{in}}}{E_\k^{\textrm{out}}} }  
\left(
\alpha_\k \,
   \rme^{-\rmi E_\k^{\textrm{out}} x^0 + \rmi\k\cdot\x}
+
\beta^*_\k \,
   \rme^{\rmi E_\k^{\textrm{out}} x^0 + \rmi\k\cdot\x}
\right) \;.
\end{equation}
By noticing that $[\phi^+_{\rmin\k}(x^0,-\x)]^*$ satisfies both the
same initial condition as $\phi^-_{\rmin\k}(x)$ and the same equation
of motion, one finds the solution at $x^0\to\infty$ that starts as a
negative energy wave as
\begin{equation}\label{eq:phiinminus}
\lim_{x^0\to+\infty}\phi^-_{\rmin\k}(x) =
\sqrt{ \frac{E_\k^{\textrm{in}}}{E_\k^{\textrm{out}}} }  
\left(
\alpha^*_\k \,
   \rme^{\rmi E_\k^{\textrm{out}} x^0 + \rmi\k\cdot\x}
+ 
\beta_\k \,
   \rme^{-\rmi E_\k^{\textrm{out}} x^0 + \rmi\k\cdot\x}
\right) \;.
\end{equation}
This can also been seen using 
$\rme^{-\rmi\k \cdot \x}\phi^+_{\rmin\k}(x) = 
[\rme^{-\rmi\k \cdot \x}\phi^-_{\rmin\k}(x)]^*.$
We can now deduce the relation,
\begin{align}\label{eq:phiinout1}
\phi^+_{\rmin\k}(x) &=
\sqrt{ \frac{E_\k^{\textrm{in}}}{ E_\k^{\textrm{out}}} }
\left(
\alpha_\k\, \phi^+_{\rmout\k}(x)
+\beta^*_\k\, \phi^-_{\rmout\k}(x)
\right) \;,
\\ \label{eq:phiinout2}
\phi^-_{\rmin\k}(x) &=
\sqrt{ \frac{E_\k^{\textrm{in}}}{ E_\k^{\textrm{out}}} }
\left(
 \alpha^*_\k\, \phi^-_{\rmout\k}(x) 
+
\beta_\k\, \phi^+_{\rmout\k}(x)
\right) \;.
\end{align}

In the general case of a space dependent background field the
Bogoliubov coefficients are not diagonal in momentum space.  However,
one can diagonalize the transformation matrix from the in- to the
out-states, and our following discussion will equally well apply to
the eigenstates of the more general transformation instead of
individual momentum modes.  Inserting the decompositions
(\ref{eq:phiinout1}) and (\ref{eq:phiinout2}) into
\eq(\ref{eq:fidecomp}) one gets
\begin{equation}
 \begin{split}
a_{\rmout\k} &= \alpha_\k\, a_{\rmin\k} + \beta_\k\, b^\dag_{\rmin-\k} \;,
\\
b_{\rmout\k}^\dag &= \alpha_{-\k}^*\, b_{\rmin\k}^\dag
 + \beta_{-\k}^*\, a_{\rmin-\k} \;.
 \end{split}
\label{eq:transform}
\end{equation}
Consistency with the commutation relations (\ref{eq:abcommut}) gives
the normalization condition,
\begin{equation}
|\alpha_\k|^2 - |\beta_\k|^2 = 1 \;.
\label{eq:normalization}
\end{equation}
This normalization condition is a consequence of the charge
conservation symmetry of our Lagrangian\footnote{One can verify that
  eq.~(\ref{eq:normalization}) implies
\begin{equation*}
{\bs Q}_{\rm out}={\bs Q}_{\rm in}\; ,\quad\mbox{where\ \ }
{\bs Q}
\equiv e\int \frac{d^3\k}{(2\pi)^3}\; (a^\dagger_\k a_\k-b^\dagger_\k b_\k)\; .
\end{equation*}}.
The above relations can be inverted to give
\begin{align}
a_{\rmin\k} &= \alpha^*_\k\, a_{\rmout\k} 
- \beta_\k\, b^\dag_{\rmout-\k} \;,
\\
b_{\rmin\k} &= \alpha^*_{-\k}\, b_{\rmout\k} 
- \beta_{-\k}\, a^\dag_{\rmout-\k} \;.
\end{align}

The vacuum is defined as the state that is annihilated by all the
destruction operators (we shall drop the in- and out-labels for a
moment),
\begin{equation}
a_\k |0\rangle = b_\k |0\rangle = 0 \;.
\end{equation}
A properly normalized state with $n$ particles of momentum $\k$
can be constructed as
\begin{equation}
|n_\k\rangle = \biggl( \frac{a^\dag_\k}{\sqrt{n! V}} \biggr)^n
 |0\rangle \;.
\end{equation}
Particles of momentum $\k$ are counted with the particle number
operator,
\begin{equation}
\frac{\rmd \hat{N}}{\rmd^3\k}= \frac{a^\dag_\k a_\k}{(2\pi)^3} \;.
\end{equation}
For example, the expectation value of the number operator on a state
with one particle of momentum $\p$ is
\begin{equation}
\langle 1_\p | \frac{\rmd \hat{N}}{\rmd^3\k} | 1_\p\rangle
=
\langle 0 | \frac{a_\p}{\sqrt{V}} \frac{a^\dag_\k a_\k}{(2\pi)^3}
\frac{a^\dag_\p}{\sqrt{V}} | 0 \rangle
= \delta(\p - \k) \;.
\end{equation}

After setting up these conventions let us return to the problem of
particle production.  We consider the situation where there are no
particles in at $x^0\to-\infty$, then the system is in the incoming
vacuum state defined by
\begin{equation}
 a_{\rmin\p} |0_{\rm in}\rangle = b_{\rmin\p}
|0_{\rm in}\rangle = 0 \;.
\end{equation}
We are working in the Heisenberg picture where there is no time
evolution in the states, so the system stays in the
$|0_{\rm in}\rangle $ state.  But at late time $x^0\to+\infty$
particles are described by the ``out'' annihilation operators which do not
necessarily annihilate $|0_{\rm in}\rangle$.  In order to count the
number of outgoing particles we must count the number of
out-particles contained in the state $|0_{\rm in}\rangle$.  For this
it is useful to derive an expression for $|0_{\rm in}\rangle$ in terms
of the ``out'' quantities $a_{\rmout\p}$, $b_{\rmout\p}$ and
$|0_{\rm out}\rangle$.  That is, we have to solve,
\begin{equation}\label{eq:tildeinvac}
 \begin{split}
 a_{\rmin\p} |0_{\rm in}\rangle
&=  \bigl( \alpha_\p^*\, a_{\rmout\p} - \beta_\p b_{\rmout-\p}^\dag \bigr)
  |0_{\rm in}\rangle = 0 \;,
\\ 
 b_{\rmin\p} |0_{\rm in}\rangle
&=  \bigl(\alpha^*_{-\p}\, b_{\rmout\p} - \beta_{-\p} a_{\rmout-\p}^\dag \bigr)
  |0_{\rm in}\rangle = 0 \,.
 \end{split}
\end{equation}
This above equation can be easily solved using the following ansatz;
\begin{equation}\label{eq:ansatz}
 |0_{\rm in}\rangle
  = C\prod_\k \exp\Bigl( \lambda_\k\, a_{\rmout\k}^\dag \,
  b_{\rmout-\k}^\dag \Bigr) |0_{\rm out}\rangle \;,
\end{equation}
where $C$ is a normalization constant.  Applying the
condition~(\ref{eq:tildeinvac}) to the ansatz~(\ref{eq:ansatz}) we
find the condition,
\begin{eqnarray}
&&\Big[\sum_{n=1}^{\infty}\frac{\lambda_\k^n}{n!} \alpha_\k^*\, n
 V (a_{\rmout\k}^\dag)^{n-1}  ( b_{\rmout-\k}^\dag)^n
\nonumber\\
&&
\qquad\qquad
-
\sum_{n=0}^\infty \frac{\lambda_\k^n}{n!} \beta_\k\,
 (a_{\rmout\k}^\dagger)^n (b_{\rmout-\k}^\dag)^{n+1} \Big]|0_{\rm out}\rangle = 0 \;.
\end{eqnarray}
Shifting the summation variable by one in the first term gives
\begin{equation}
\sum_{n=0}^\infty
\frac{\lambda_\k^n}{n!} \bigl[ \lambda_\k \alpha_\k^*  V
-
\beta_\k \bigr] ( a_{\rmout\k}^\dag)^n ( b_{\rmout-\k}^\dag)^{n+1}
 |0_{\rm out}\rangle = 0 \;,
\end{equation}
which is satisfied by
\begin{equation}
 \lambda_\p = V^{-1} \,\frac{\beta_\p}{\alpha_\p^\ast} \;.
\end{equation}
Now we can fix the normalization constant from
\begin{equation}
 \begin{split}
\langle 0_{\rm out}|
\exp \Bigl\{\lambda_\p^\ast a_{\rmout\p}\, b_{\rmout-\p} \Bigr\} \,
\exp\Bigl\{\lambda_\p a_{\rmout\p}^\dag\, b_{\rmout-\p}^\dag \Bigr\}
|0_{\rm out}\rangle \\
 = \frac{1}{1-V^2|\lambda_\p|^2} = 1 + |\beta_\p|^2 \;,
 \end{split}
\end{equation}
where we used the normalization condition~\nr{eq:normalization}.  We
then have the expression for the initial state in the following form:
\begin{equation}\label{eq:instate}
 |0_{\rm in}\rangle = \prod_\k (1+|\beta_\k|^2)^{-1/2} \exp\biggl[
  V^{-1}\,\frac{\beta_\k}{\alpha_\k^\ast}\; a_{\rmout\k}^\dag\,
  b_{\rmout-\k}^\dag \biggr] |0_{\rm out}\rangle \;.
\end{equation}
This has a clear physical interpretation that the initial vacuum is
a superposition of states with out-state pairs of particles with $\k$
and  antiparticles with $-\k$.

Armed with the explicit expression~(\ref{eq:instate}) it is now
straightforward to calculate, for example, single and double inclusive
spectra by taking expectation values of the number operator.  The
spectrum of particles is
\begin{equation}
 \frac{\rmd N_1^+}{\rmd^3\p} \!=\! \langle 0_{\rm in}| 
\frac{a_{\rmout\p}^\dag a_{\rmout\p}}{(2\pi)^3}
 |0_{\rm in}\rangle
  \!=\! \frac{V}{(2\pi)^3} (1+|\beta_\p|^2)^{-1} \sum_{n=1}^\infty
  n \biggl( \frac{|\beta_\p|^2}{|\alpha_\p|^2} \biggr)^n
\!=\! \frac{V}{(2\pi)^3} |\beta_\p|^2
\;.
\end{equation}
This expression exactly coincides with \eq(\ref{eq:1+}).
The two particle spectrum (equal sign) is likewise
\begin{align}
 \!\frac{\rmd N_2^{++}}{\rmd^3\p_1 \rmd^3\p_2} &=
\langle 0_{\rm in}|
\left(
\frac{a_{\rmout\p_1}^\dag a_{\rmout\p_1}}{(2\pi)^3}
\frac{a_{\rmout\p_2}^\dag a_{\rmout\p_2}}{(2\pi)^3}
\!-\! \delta(\p_1-\p_2) \frac{a_{\rmout\p_1}^\dag a_{\rmout\p_1}}{(2\pi)^3}
\right)
 |0_{\rm in}\rangle \notag\\
&=
\frac{1}{(2\pi)^6}\;
\langle 0_{\rm in}|
a_{\rmout\p_1}^\dag a_{\rmout\p_2}^\dag a_{\rmout\p_1} a_{\rmout\p_2}
 |0_{\rm in}\rangle \notag\\
&= \frac{\rmd N_1^+}{\rmd^3\p_1}\; \frac{\rmd N_1^+}{\rmd^3\p_2}
 +
\delta(\p_1-\p_2) \frac{V}{(2\pi)^3} |\beta_{\p_1}|^4 \:.
\end{align}
Note that we are explicitly subtracting the delta function
contribution to agree with the definition~(\ref{eq:N2++int}).  The
variance is
\begin{equation}
\big<{\bs N}^+{\bs N}^+\big> - \big<{\bs N}^+\big>\big< {\bs N}^+\big> =
\int \rmd^3\p\; n_\p(1+f_\p) \;,
\end{equation}
which should remind the reader of the particle number fluctuations in
a Bose-Einstein system.  It is actually easy to see that this is
indeed what we have by looking directly at the probability
distribution.

In order to see more clearly the structure of the probability
distribution we note that we can decompose the Fock space into a
direct product of the Fock spaces for different momenta $\k$.  We then
write an arbitrary state $|\Psi \rangle$ as a tensor product
\begin{equation}
|\Psi \rangle = \bigotimes_\k |\Psi \rangle_\k \;.
\end{equation}
It is convenient to group together particles of momentum $\k$ and
antiparticles of momentum $-\k$ under the same label $\k$.  Thus we
define the vacuum states $|0,0\rangle_\k$ of the subspaces $\k$ as
\begin{equation}
a_{\rmout\k} |0,0\rangle_\k = b_{\rmout-\k} |0,0\rangle_\k = 0 \;.
\end{equation}
We can now write $|0_{\rm out}\rangle$ as a tensor product of the
vacuum states of the different modes,
\begin{equation}
|0_{\rm out}\rangle = \bigotimes_\k |0,0\rangle_\k \;.
\end{equation}
A state with $m$ particles of momentum $\k$ and $n$ antiparticles of
momentum $-\k$ at $x^0 \to \infty$ is then denoted by
\begin{equation}
|m,n\rangle_\k = \left(\frac{a_{\rmout\k}^\dag}{\sqrt{m!V}}\right)^m
\left(\frac{b_{\rmout-\k}^\dag}{\sqrt{n!V}}\right)^n |0,0\rangle_\k \;.
\end{equation}
By taking tensor products of the states $|m,n\rangle_\k$ for different
$\k$ one can construct a complete basis for the Fock space.

Applying this decomposition to \eq(\ref{eq:instate}) we write
\begin{equation}
|0_{\rm in}\rangle = \bigotimes_\k
\left\{
 (1+|\beta_\k|^2)^{-1/2} \exp\biggl[
   V^{-1}\frac{\beta_\k}{\alpha_\k^\ast} a_{\rmout\k}^\dag\,
   b_{\rmout -\k}^\dag \biggr] |0,0\rangle_\k
\right\} \;,
\end{equation}
and expanding the exponential we get
\begin{align}
|0_{\rm in}\rangle 
&= \bigotimes_\k
\left\{
 (1+|\beta_\k|^2)^{-1/2}
 \sum_{m_\k =0}^\infty
 \frac{1}{m_\k !}
 \left( V^{-1}\frac{\beta_\k}{\alpha_\k^\ast} a_{\rmout\k}^\dag\,
   b_{\rmout -\k}^\dag \right)^{m_\k}
 |0,0\rangle_\k
\right\}
\\
& =
 \bigotimes_\k
\left\{
 (1+|\beta_\k|^2)^{-1/2}
 \sum_{m_\k =0}^\infty
 \left(\frac{\beta_\k}{\alpha_\k^\ast} \right)^{m_\k}
 |m_\k,m_\k\rangle_\k
\right\} \;.
\end{align}
Thus we see that explicitly the ``in'' vacuum for a momentum mode $\k$
is a superposition of outgoing particle-antiparticle pair states.  The
amplitude for being in a state with $m_\k$ pairs is
\begin{equation}\label{eq:kampli}
\mathcal{M}_\k =
 {}_{\k}\langle m_\k, m_\k| \mathbb{P}_\k | 0_{\rm in} \rangle =
 \frac{1}{ \sqrt{1+|\beta_\k|^2} }
 \left(\frac{\beta_\k}{\alpha^*_\k}\right)^{m_\k} \;,
\end{equation}
where we have to introduce $\mathbb{P}_\k$, the projection operator to
the subspace $\k$, to take the inner product\footnote{The states
$|0_{\rm in}\rangle$ and $|m_\k,m_\k\rangle_\k$ live in different
  Hilbert spaces; one in the whole Fock space and the other one in its
  subspace $\k$.  In order to take an inner product one therefore has
  to project out the $\k$ component of the state $|0_{\rm in}\rangle$.
  Physically this projection means that we are not measuring the other
  momentum modes of the state $|0_{\rm in}\rangle$ than $\k$.  Thus
  \eq(\ref{eq:kampli}) gives the amplitude to have $m_\k$ pairs in the
  mode $\k$ and \emph{any number} of particles in the other momentum
  states.}.  The corresponding probability to have $m_\k$ pairs in the
mode $\k$ and any number of particles in the other momentum modes is
\begin{equation}
 \begin{split}
P(m_\k) &= \Bigl| {}_\k\langle m_\k, m_\k | \mathbb{P}_\k
 |0_{\rm in}\rangle \Bigr|^2 \\
&=  \frac{1}{1+|\beta_\k|^2 }
 \left( \frac{|\beta_\k|^2}{1+|\beta_\k|^2 }\right)^{m_\k}
=  \frac{1}{1+f_\k } \left( \frac{f_\k}{1+f_\k }\right)^{m_\k} \;.
 \end{split}
\end{equation}
This law for the probabilities characterizes the Bose-Einstein, or
geometrical, distribution.  Since the momentum modes are independent,
we can then write the combined probability distribution as
\begin{equation}
P(\{m_\k\}) = \prod_\k P(m_\k) =
\prod_\k \frac{1}{1+f_\k}
  \left( \frac{f_\k}{1+f_\k}\right)^{m_\k} \;,
\label{eq:probdist2}
\end{equation}
which is exactly the form at which we already arrived in
\eq(\ref{eq:probdist}).

\section{Exactly solvable example}
\label{sec:exact}
So far our discussions and formulas are given in a rather general way.
In what follows, we consider an example which is exactly solvable and
demonstrate how the formulas in the preceding sections lead to the
concrete evaluation of spectrum of produced particle under an external
electric field.

The LSZ reduction method and the interpretation as a Bogoliubov
transformation both require that the asymptotic states at
$x^0\to\pm\infty$ are well defined.  This means that the external
fields should be adiabatically vanishing so that we can define
$|0_{\rm in}\rangle$ and $|0_{\rm out}\rangle$ without ambiguity.
Thus, the imposed electric fields must be time-dependent, beginning
with zero at $x^0=-\infty$, growing finite with increasing $x^0$, and
diminishing to zero again as $x^0\to+\infty$.

\subsection{Choice of the gauge potential}
\begin{figure}[htbp]
\begin{center}
\vspace*{5mm}
\includegraphics[width=0.7\textwidth]{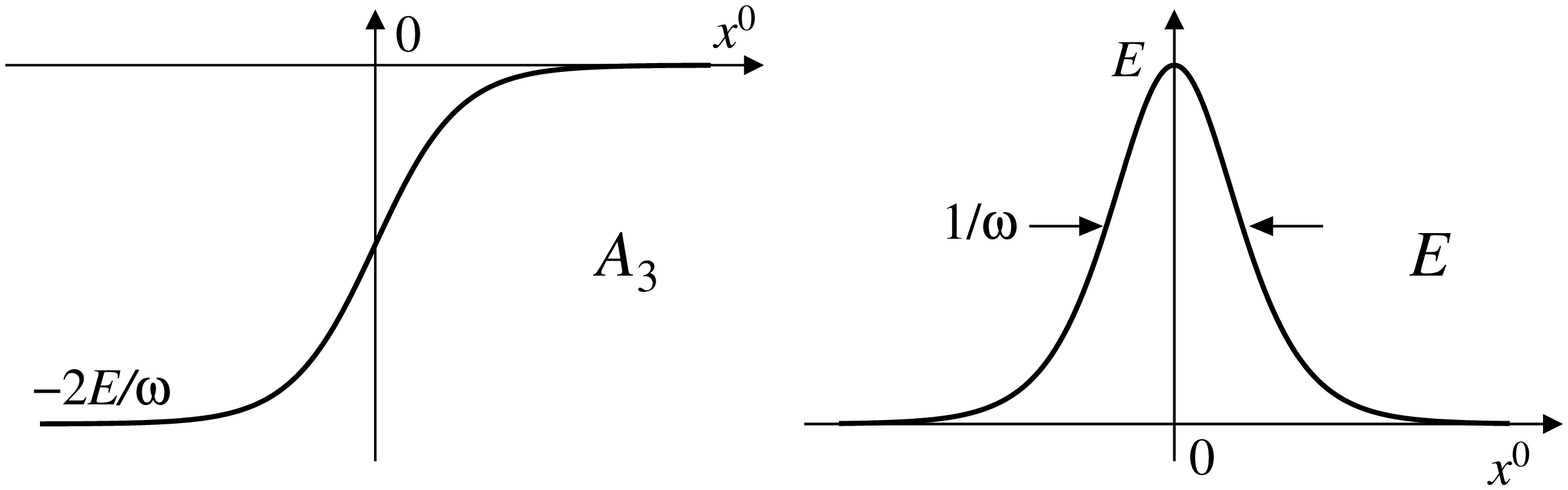}
\end{center}
\caption{\label{fig:a3_e}Sketch of the chosen gauge potential and the
  associated electric field as a function of time $x^0$.  The electric
  field has a peak at $x^0=0$ whose height is $E$ and width is
  specified by $1/\omega$.  The gauge potential has an offset from
  zero by $-2E/\omega$ in the in-vacuum at $x^0=-\infty$, meaning that
  the origin of $p_3$ is shifted by this offset.}
\end{figure}

It is known that the Klein-Gordon or Dirac equation under the
following time-dependent electric field $\bE=(0,0,E(x^0))$ is
exactly solvable~\cite{Dunne:2004nc,Nikishov:1970br,Hebenstreit:2008ae};
\begin{equation}
 E(x^0) = \frac{E}{\bigl[ \cosh(\omega x^0) \bigr]^2} \;.
\label{eq:E}
\end{equation}
This electric field exponentially goes to zero for
$|x^0|\gg \omega^{-1}$.  In the limit of $\omega\to0$ the electric
field becomes homogeneous in time.  We can choose a gauge in which
$A_0=0$ and the vector potential associated with the electric field
(\ref{eq:E}) is $\bA=-(0,0,A_3=\int\!\rmd x^0 E(x^0))$\footnote{We
  work with the $(+,-,-,-)$ metric convention.}.  
We can
immediately recover \eq(\ref{eq:E}) from
$\bE=-\boldsymbol{\nabla}A^0-\partial_0 \bA$.  After the integration
we find the Sauter-type gauge potential,
\begin{equation}
 A_3(x) = \int^{x^0} \!\rmd y^0 E(y^0)
  = \frac{E}{\omega}\bigl[\tanh(\omega x^0)-1 \bigr] \;,
\end{equation}
where we have chosen the integration constant so as to make
$A_3(x)\to0$ at $x^0\to+\infty$.  As we have discussed in
\se\ref{sec:bogo} this is a very natural choice in the present case.
A constant $A_3$ amounts to a shift in the third component of the
momentum $p_3$, which is interpreted as a different frame choice.  It
is most natural to sit in the frame in which the particles and
antiparticles measured at $x^0=+\infty$ are at rest if their $p_3$ is
zero.  The gauge potential and the electric field are sketched in
\fig\ref{fig:a3_e}.

\subsection{Solving the equation of motion}
An explicit solution for the Sauter-type gauge potential is already
known, and so we will simply explain the necessary notation and then
jump into the known expression of the solution.  To make this paper as
self-contained as possible we supplement the derivation in
appendix~\ref{app:solve} in more detail. Introducing the following
notation,
\begin{equation}
 \lambda \equiv \frac{eE}{\omega^2} \;,
\label{eq:lambda}
\end{equation}
the equation of motion in scalar QED is given by the gauged
Klein-Gordon equation as
\begin{equation}
 \Bigl[ \partial_0^2 - \bigl(\partial_3 -\rmi\lambda\omega
  \bigl[\tanh(\omega x^0)-1\bigr]\bigr)^2 -\partial_\perp^2
  +m^2 \Bigr]\phi(x) = 0 \;,
\end{equation}
which is immediately read from the Lagrangian
density~(\ref{eq:lagrangian}).  Because the equation of motion does
not have explicit dependence on $\x$, we can factorize the wave
function $\phi(x)$ into the time-dependent part and the spatial
plane-wave part, i.e.\
\begin{equation}\label{eq:psifact}
 \phi(x) = \psi_\k(x^0)\;\rme^{\rmi \k\cdot\x} \;.
\end{equation}
The time dependence is governed by the differential equation,
\begin{equation}\label{eq:keom}
 \Bigl[ \partial_0^2 + \bigl( k_3 + \lambda\omega\bigl[
  \tanh(\omega x^0)-1\bigr]\bigr)^2 + k_\perp^2 + m^2 \Bigr]
  \psi_\k(x^0) = 0 \;.
\end{equation}
We here change the variable $x^0$ into $\xi$, defined by
\begin{equation}
 \xi \equiv \frac{1}{2}\bigl[ \tanh(\omega x^0) + 1 \bigr] \;.
\end{equation}
With this variable we can eliminate the hyperbolic function from the
equation and express it only in terms of meromorphic functions.
Moreover, $\xi$ has a convenient asymptotic behavior. We will later
make use of
\begin{equation}
 \xi^{\rmi \alpha} = \biggl( \frac{\rme^{2\omega x^0}}
  {1+\rme^{2\omega x^0}} \biggr)^{\rmi \alpha} \longrightarrow
  \left\{ \begin{array}{lp{.1em}l}
   1 && \mbox{for $x^0\to+\infty$} \\
   \rme^{2\rmi\alpha\omega x^0} && \mbox{for $x^0\to-\infty$}
  \end{array} \right.
\label{eq:asym1}
\end{equation}
and also
\begin{equation}
 (1-\xi)^{\rmi \alpha} = \biggl( \frac{\rme^{-2\omega x^0}}
  {1+\rme^{-2\omega x^0}} \biggr)^{\rmi \alpha} \longrightarrow
  \left\{ \begin{array}{lp{.1em}l}
   \rme^{-2\rmi\alpha\omega x^0} && \mbox{for $x^0\to+\infty$} \\
   1 && \mbox{for $x^0\to-\infty$}
 \end{array} \right.
\label{eq:asym2}
\end{equation}
to infer the plane-wave boundary conditions.

Also, for concise notation we introduce the dimensionless energies in
the same way as in Dunne's review~\cite{Dunne:2004nc};
\begin{equation}
 \mu \equiv \frac{E_\k^{\textrm{in}}}{2\omega} \;,\qquad
 \nu \equiv \frac{E_\k^{\textrm{out}}}{2\omega} \;,
\end{equation}
where, with the time-dependent gauge potential, the energies in the
in- and out-states are, respectively,
\begin{equation}
 (E_\k^{\textrm{in}})^2 = (k_3 - 2\lambda\omega)^2 + \k_\perp^2
  + m^2 \,,\qquad
 (E_\k^{\textrm{out}})^2 = k_3^2 + \k_\perp^2 + m^2 \;.
\end{equation}
We note that the energy of antiparticles is given by
$E_{-\k}^{\textrm{in}}$ in the in-vacuum and $E_{-\k}^{\textrm{out}}$
in the out-vacuum, respectively.  Also we should explain our
convention of the electric charge $e$.  Our choice is as follows;
particles are \emph{negatively} charged and antiparticles are
\emph{positively} charged for $e>0$.  These are reminiscent of
electrons and positrons in real QED.\ \ Therefore, in the above,
noticing that $k^3=-k_3$ we see that the particle dispersion relation
$E_\k^{\textrm{in}}$ starts with a larger longitudinal momentum than
$E_\k^{\textrm{out}}$, which is understood as the deceleration by the
Lorentz force in the direction anti-parallel to the external electric
field.

Now we are ready to express the solution of the differential equation.
One can express two linearly independent solutions in terms of the
hypergeometric functions as
\begin{equation}
 \begin{split}
 \psi_\k^{(1)}(\xi) &= \xi^{-\rmi\mu}(1\!-\!\xi)^{-\rmi\nu}
  \Hyg{\half\!-\!\rmi(\lambda'\!+\!\mu\!+\!\nu)}
  {\half\!+\!\rmi(\lambda'\!-\!\mu\!-\!\nu)}
  {1\!-\!2\rmi\mu}{\xi} ,\\
 \psi_\k^{(2)}(\xi) &= \xi^{\rmi\mu}(1-\xi)^{-\rmi\nu}
  \Hyg{\half\!-\!\rmi(\lambda'\!-\!\mu\!+\!\nu)}
  {\half\!+\!\rmi(\lambda'\!+\!\mu\!-\!\nu)}
  {1\!+\!2\rmi\mu}{\xi} \;,
 \end{split}
\label{eq:hyper}
\end{equation}
where $\lambda'\equiv\sqrt{\lambda^2-1/4}$ with $\lambda$ defined in
\eq(\ref{eq:lambda}).

\subsection{Asymptotic behavior of the solution} 
Now we need to take an appropriate linear combination of the two
solutions written in eq.~(\ref{eq:hyper}), so that the the boundary
condition like \eq(\ref{eq:EOM-BC}) can be fulfilled.  In fact, it
will turn out that these solutions already satisfy a simple plane-wave
boundary condition.  To make this explicit we should use
\eqs(\ref{eq:asym1}) and (\ref{eq:asym2}) together with the general
property of the hypergeometric function; $\Hyg{a}{b}{c}{\xi\to0}\to1$.
Therefore, in the limit of $x^0\to-\infty$ (i.e.\ $\xi\to0$), we see,
\begin{equation}
 \psi_\k^{(1)}(\xi\to0) \to \rme^{-2\rmi\mu\omega x^0}
 = \rme^{-\rmi E_\k^{\textrm{in}}x^0} \;,
 \quad
 \psi_\k^{(2)}(\xi\to0) \to \rme^{2\rmi\mu\omega x^0}
 = \rme^{\rmi E_\k^{\textrm{in}}x^0} \;.
\end{equation}
In accord with the identification defined by \eqs(\ref{eq:planewave1})
and (\ref{eq:planewave2}) we have
\begin{equation}
 \phi_{\rmin\k}^+ = \psi_\k^{(1)} \,\rme^{\rmi \k\cdot\x} \;,
 \qquad
 \phi_{\rmin \k}^- = \psi_\k^{(2)} \,\rme^{\rmi \k\cdot\x} \;,
\end{equation}

The quantities necessary to compute the particle and antiparticle
production are obtained as the coefficient of these solutions in the
$x^0\to+\infty$ limit--when decomposed in terms of the
$\phi^\pm_{\rmout \k}$.  To this end, it is necessary to know the limit
of $\Hyg{a}{b}{c}{\xi\to1}$.  One must be, however, very careful when
taking this limit because $a$, $b$, and $c$ are complex numbers in
this case.  Thus, it is convenient to make a transformation in the
argument from $\xi$ to $1-\xi$, which is possible by means of the
following mathematical identity;
\begin{equation}
 \begin{split}
 & \Hyg{a}{b}{c}{x} = \frac{\Gamma(c)\Gamma(c-a-b)}
  {\Gamma(c-a)\Gamma(c-b)}\, \Hyg{a}{b}{1-c+a+b}{1-x} \\
 &\quad + \frac{\Gamma(c)\Gamma(a\!+\!b\!-\!c)}
  {\Gamma(a)\Gamma(b)}\, (1-x)^{c-a-b}
  \Hyg{c-a}{c-b}{1\!+\!c\!-\!a\!-\!b}{1-x} \,,
 \end{split}
\label{eq:identity}
\end{equation}
which leads to an alternative expression of the solution,
\begin{align}
  & \psi_\k^{(1)}(\xi) = \xi^{-\rmi\mu}(1\!-\!\xi)^{-\rmi\nu}
  B_\k^\ast\, \Hyg{\half\!-\!\rmi(\lambda'\!+\!\mu\!+\!\nu)}
  {\half\!+\!\rmi(\lambda'\!-\!\mu\!-\!\nu)}
  {1\!-\!2\rmi\nu}{1\!-\!\xi} \notag\\
 &\; + \xi^{-\rmi\mu}(1\!-\!\xi)^{\rmi\nu} A_\k\,
  \Hyg{\half\!+\!\rmi(\lambda'\!-\!\mu\!+\!\nu)}
  {\half\!-\!\rmi(\lambda'\!+\!\mu\!-\!\nu)}
  {1\!+\!2\rmi\nu}{1\!-\!\xi} \;,
\end{align}
and
\begin{align}
  & \psi_\k^{(2)}(\xi) = \xi^{\rmi\mu}(1\!-\!\xi)^{-\rmi\nu} A_\k^\ast\,
  \Hyg{\half\!-\!\rmi(\lambda'\!-\!\mu\!+\!\nu)}
  {\half\!+\!\rmi(\lambda'\!+\!\mu\!-\!\nu)}
  {1\!-\!2\rmi\nu}{1\!-\!\xi} \notag\\
 &\; + \xi^{\rmi\mu}(1\!-\!\xi)^{\rmi\nu} B_\k\,
  \Hyg{\half\!+\!\rmi(\lambda'\!+\!\mu\!+\!\nu)}
  {\half\!-\!\rmi(\lambda'\!-\!\mu\!-\!\nu)}
  {1\!+\!2\rmi\nu}{1\!-\!\xi} \;,
\end{align}
where we defined,
\begin{equation}
 \begin{split}
 A_\k &\equiv \frac{\Gamma(1-2\rmi\mu)\Gamma(-2\rmi\nu)}
  {\Gamma(\half-\rmi(\lambda'+\mu+\nu))
  \Gamma(\half+\rmi(\lambda'-\mu-\nu))} \;,\\
 B_\k^\ast &\equiv \frac{\Gamma(1-2\rmi\mu)\Gamma(2\rmi\nu)}
  {\Gamma(\half+\rmi(\lambda'-\mu+\nu))\Gamma(\half-\rmi(\lambda'+\mu-\nu))}
 \;,
 \end{split}
\end{equation}

Here we note that, if we take a naive limit of $x\to1$ using a
textbook formula on $\Hyg{a}{b}{c}{1}$, it would miss the second term
in the identity~(\ref{eq:identity}) because of the factor 
$(1-x)^{c-a-b}$ which is zero if $a$, $b$, $c$ are real but is an
oscillatory finite function if $a$, $b$, $c$ are complex.  At this
point it is straightforward to deduce the asymptotic plane-wave forms
at $x^0\to\infty$, that is, $\xi\to1$;
\begin{equation}
 \begin{split}
 & \psi_\k^{(1)}(\xi\to1)\to A_\k\, \rme^{-2\rmi\nu\omega x^0}
  + B_\k^\ast\, \rme^{2\rmi\nu\omega x^0}
  = A_\k\, \rme^{-\rmi E_\k^{\textrm{out}}x^0}
  + B_\k^\ast\, \rme^{\rmi E_\k^{\textrm{out}}x^0} \;,\\
 & \psi_\k^{(2)}(\xi\to1)\to A_\k^\ast\, \rme^{2\rmi\nu\omega x^0}
  + B_\k\, \rme^{-2\rmi\nu\omega x^0}
  = A_\k^\ast\, \rme^{\rmi E_\k^{\textrm{out}}x^0}
  + B_\k\, \rme^{-\rmi E_\k^{\textrm{out}}x^0} \;.
 \end{split}
\end{equation}
Comparing the above behavior with the Bogoliubov
transformations~(\ref{eq:phiinplus}) and (\ref{eq:phiinminus}), we
obtain the Bogoliubov coefficients as
\begin{equation}
 \alpha_\k = \sqrt{\frac{E_\k^{\textrm{out}}}{E_\k^{\textrm{in}}}}
  A_\k \;,\qquad
 \beta_\k = \sqrt{\frac{E_\k^{\textrm{out}}}{E_\k^{\textrm{in}}}}
  B_\k \;,
\end{equation}

\subsection{Particle spectrum}
\begin{figure}[htbp]
\begin{center}
\includegraphics[width=0.7\textwidth]{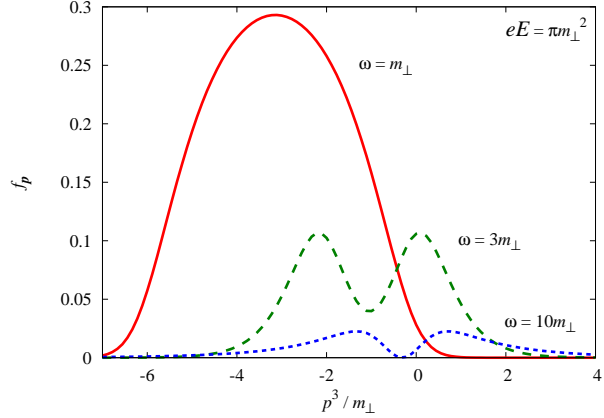}
\end{center}
\caption{\label{fig:fp}Time scale dependence of the produced particle
  distribution as a function of $p^3$.  As $\omega\to0$ the
  distribution approaches \eq(\ref{eq:wsmall}) and $f_\p$ extends
  between $p^3\simeq - 2eE/\omega$ and $p^3\simeq 0$as seen from the
  curve for $\omega=m_T$ in the figure.  In contrast, with increasing
  $\omega$, the result approaches \eq(\ref{eq:wlarge}) which spreads
  wider than the small-$\omega$ case with the distribution center
  located at $p^3\simeq - eE/\omega$.
}
\end{figure}

Before addressing the concrete expressions for the particle spectrum,
we list all the necessary formulas to proceed with the calculations.
The Gamma function generally satisfies,
\begin{equation}
 \Gamma(1+z) = z\Gamma(z) \,,\qquad
 \Gamma(1-z)\Gamma(z) = \frac{\pi}{\sin(\pi z)} \,,
\end{equation}
from which we can derive the following useful relations,
\begin{equation}
 |\Gamma(\rmi\alpha)|^2 = \frac{\pi}{\alpha\sinh(\pi\alpha)}
  \,,\quad
 |\Gamma(1+\rmi\alpha)|^2 = \frac{\pi\alpha}{\sinh(\pi\alpha)}
  \,,\quad
 |\Gamma(\half+\rmi\alpha)|^2 = \frac{\pi}{\cosh(\pi\alpha)} \,.
\end{equation}
Then, after some algebra, we reach,
\begin{equation}
 |\alpha_\k|^2 = \frac{\cosh[\pi(\lambda'+\mu+\nu)]\,
  \cosh[\pi(\lambda'-\mu-\nu)]}{\sinh(2\pi\mu)\,
  \sinh(2\pi\nu)} \;.
\end{equation}
and
\begin{equation}
 |\beta_\k|^2 = \frac{\cosh[\pi(\lambda'-\mu+\nu)]\,
  \cosh[\pi(\lambda'+\mu-\nu)]}{\sinh(2\pi\mu)\,
  \sinh(2\pi\nu)} \,.
\end{equation}
We note here that, using
$\cosh(a+b)=\cosh(a)\cosh(b)+\sinh(a)\sinh(b)$ twice, we can easily
check that
\begin{equation}
 |\alpha_\k|^2 - |\beta_\k|^2 = 1 \;,
\end{equation}
which is consistent with the condition~(\ref{eq:normalization}).  Now
that we have $|\beta_\k|^2$ explicitly, we can get the general
probability distribution which is characterized only in terms of
$|\beta_\k|^2$.  The single inclusive spectrum, for example, is
\begin{equation}
 \frac{\rmd N_1^+}{\rmd^3\p}
 = \frac{V}{(2\pi)^3}\, \frac{\cosh[\pi(\lambda'-\mu_\p+\nu_\p)]\,
  \cosh[\pi(\lambda'+\mu_\p-\nu_\p)]}{\sinh(2\pi\mu_\p)\,
  \sinh(2\pi\nu_\p)} \;.
\label{eq:exact_single}
\end{equation}
From this expression we can get the occupation number $f_\p$ which is
obtained by removing the volume factor $V/(2\pi)^3$ of the single
particle spectrum.  Once $f_\p$ is given, the whole probability
distribution is known as discussed in the previous sections.  We plot
$f_\p$ as a function of $p^3$ in the unit of
$m_\perp\equiv\sqrt{p_\perp^2+m^2}$ in \fig\ref{fig:fp}.  In drawing
\fig\ref{fig:fp} we fixed $m_\perp$, and set the electric field to the
value $E=\pi m_\perp^2/e$--which is sufficiently strong to create
particles in view of the standard expression of the Schwinger
mechanism--, and then we vary the time scale $\omega$.

Now let us consider two extreme cases.  First, we take the constant
field limit ($\omega\to0$), which make the above expression as simple
as follows;
\begin{equation}
 \frac{\rmd N_1^+}{\rmd^3\p}
 \to \frac{V}{(2\pi)^3} \exp\biggl[ -\frac{\pi(p_\perp^2+m^2)}
  {4eE}\Bigl( \frac{1}{1+\rho} - \frac{1}{\rho} \Bigr)
  \biggr]  \qquad (\omega\to0) \,,
\label{eq:wsmall}
\end{equation}
where $\rho\equiv \omega p^3/(2eE)$ taking a value in the range of
$-2eE/\omega<p^3<0$ (i.e.\ $-1<\rho<1$).  In the outside region, $p^3>0$
or $p^3<-2eE/\omega$, the result is zero in the $\omega\to0$ limit.

This distribution of particles in the $p^3$ direction corresponds to
the momentum distribution as a result of the deceleration by the
electric field while it is imposed.  If $\omega$ is small, the
electric field lives long, and the produced particles are pushed down
by the Lorentz force along the $p^3$ direction for longer time.  Note
that this range bounded from $-2eE/\omega$ to zero coincides with the
range of $A_3$ changing between $x^0=\pm\infty$.  One might expect
that the total number of produced particles diverges in the case of a
constant (in time) electric field.  This is indeed true; the result of
the $p^3$ integration is nearly proportional to the integration range
$2eE/\omega$ when $\omega$ is small enough, which is divergent as
$1/\omega$--i.e. as the time during which the external electric field
is non-zero.

Next, we shall take a look at the opposite limit, i.e.\ a short-pulse
limit, $\omega\to\infty$.  In this limit \eq(\ref{eq:exact_single}) is
reduced to,
\begin{equation}
 \frac{\rmd N_1^+}{\rmd^3\p}
 \to \frac{V}{(2\pi)^3}
 \frac{E_\p^{\textrm{in}}E_\p^{\textrm{out}}}{4}
  \biggl(\frac{1}{E_\p^{\textrm{in}}}-\frac{1}{E_\p^{\textrm{out}}}
  \biggr)^2
  \qquad (\omega\to\infty) \;.
\label{eq:wlarge}
\end{equation}
To have non-zero value the electric field $eE$ should be larger than
$\omega E_\p^{\textrm{out}}$.  Even though there is no exponential
suppression\footnote{When the electric field has a fast time
  dependence, the perturbative process $\gamma\to \phi\phi^*$ can
  produce particles.}, as compared to the small-$\omega$ case, the
resulting $f_\p$ is significantly suppressed by large $\omega$ for a
fixed maximal strength of the electric field, as is apparent in
\fig\ref{fig:fp}.

Let us finish this subsection with a comment on a related work by
Cooper and Nayak~\cite{Cooper:2006mt}. In this paper, the authors
study the Schwinger mechanism for the pair production of charged
scalars in the presence of an arbitrary time-dependent background
electric field.  In particular, they calculate the pair production
rate (i.e. the number of pairs created per unit of time)\footnote{One
  may question whether this is a well defined concept, since the
  presence of a time dependent external field makes the definition of
  proper particle states ambiguous. One may argue that the only 
  quantities that
  can be defined unambiguously are those where measurements are done
  only after the external field has died out. From the point of view
  of the Bogoliubov transformation, the transformation coefficients
  are uniquely determined from the asymptotic plane-wave
  time-dependence.  Of course one may use some working definition of
  the Bogoliubov coefficients at arbitrary intermediate time, but such
  a treatment implicitly assumes a quasi-static approximation.}, and
conclude that ``the result has the same functional dependence on $E$ as
the constant electric field $E$ result with the replacement: $E \to
E(t)$''.  However, the comparison of the two limits in
eqs.~(\ref{eq:wsmall}) and (\ref{eq:wlarge}) indicates that the
time-dependent case is unlikely to be given by a mere replacement $E
\to E(t)$ in the time-independent result.

\section{Conclusions}

We have calculated the multiplicity of produced particles under a
spatially constant but time-dependent electric field in scalar QED
using a formalism based on the LSZ reduction formula.  We defined and
computed the generating functional of the particle and antiparticle
distribution, from which we determined the whole distribution of the
production probability.  We found that particle production in one
momentum mode follows a Bose-Einstein law, reflecting the statistics
of scalar particles.  We have also derived the same results by means
of a Bogoliubov transformation on creation and annihilation operators.

A natural question may arise as follows; what is the distribution not
in spinor (ordinary) QED instead of scalar QED?  Our discussions in
this paper were quite simple because we focused only on spin-zero
scalar particles.  In the case of spinor particles we need to deal
with the spin structure, which brings unessential complication in,
though the generalization is straightforward.  We shall here just
mention that the distribution of fermionic particles in the same
momentum and spin mode follows a Fermi-Dirac distribution.  We can
confirm this immediately by replacing the commutation relation by the
anticommutation one in the derivation based on the Bogoliubov
transformation.  The transformation~\nr{eq:transform} is unchanged
since this originates from the asymptotic behavior of the
wavefunctions.  The normalization condition~\nr{eq:normalization}
should be $|\alpha_\k|^2+|\beta_\k|^2=1$ then to preserve the
anticommutation relation of the transformed operators.  It is easy to
show that $|\beta_\p|^2$ gives the occupation number $f_\p$, and using
the above normalization condition we can arrive at the probability
distribution,
\begin{equation}
 P(\{m_{s,\k}\}) = \prod_{s,\k} \bigl( 1-f_{s,\k} \bigr)
  \left( \frac{f_{s,\k}}{1-f_{s,\k}} \right)^{m_{s,\k}} \;,
\label{eq:probdist3}
\end{equation}
where $s$ refers to the spin and $m_{s,\k}$ takes the values $0$ or
$1$.  Equation~(\ref{eq:probdist}) (or equivalently
(\ref{eq:probdist2})) and the above (\ref{eq:probdist3}) are our
central results.  We see that, if we drop higher orders than the
quadratic terms in $f_\k$ for $f_\k\ll 1$, all of the Bose-Einstein,
Fermi-Dirac, and Poisson distributions are trivially reduced to
identical answer; $1-f_\k$ for no-particle production and $f_\k$ for
one-particle production.  The difference emerges at the quadratic
order, and it is notable that the difference remains no matter how
small $f_\k$ is.  For instance, the two-particle production
probability in the same mode is $f_\k^2/2$ if the distribution is a
Poisson one, $f_\k^2$ if a Bose-Einstein one, and zero if a
Fermi-Dirac one.  In other words one must properly take account of the
quantum statistical nature if the multiparticle correlations are
concerned.

In the final section of this paper, we have revisited a known exactly
solvable example of time-dependent electric fields.  The
time-dependence of the Sauter-type potential is actually ideal to
think of the Schwinger mechanism; since we can unambiguously define
the asymptotic states in the infinite past and future.  This property
of adiabatically vanishing external fields is necessary to make the
discussion of particle production meaningful.  

Using the exact solution we took two extreme limits of constant
($\omega\to0$) and short-pulse ($\omega\to\infty$) electric fields.
In the constant case we found that $f_\p$ distributes almost uniformly
over the range $-2eE/\omega<p^3<0$.  Therefore the total (integrated)
number of produced particles diverges as $1/\omega$ that is
interpreted as the time duration for which the external electric field
is imposed.  In the short-pulse case, on the other hand, $f_\p$ is a
double-peak structure and the minimum in-between is located at
$p^3=-eE/\omega$.  In practice the latter would be useful because it
is more difficult to sustain larger $eE/\omega$ in the laboratory.

\section*{Acknowledgments}
We thank Raju Venugopalan for discussions.  K.~F.\ thanks Harmen
Warringa for discussions on the time-dependent electric field.
K.~F.\ is grateful to Institut de Physique Th\'{e}orique
CEA/DSM/Saclay for warm hospitality where this work was initiated and
he is supported by Japanese MEXT grant No.\ 20740134 and also
supported in part by Yukawa International Program for Quark Hadron
Sciences. T.~L.\ is supported by the Academy of Finland, project
126604.  F.~G.\ is supported in part by Agence Nationale de la
Recherche via the programme ANR-06-BLAN-0285-01.

\appendix

\section{Multiplicity distribution}
\label{app:multiplicity}
In some cases, one is interested only in the overall number of
particles and antiparticles in the final state, but not in their
distribution in momentum space.  For such observables, one can define
a simpler generating function that does not contain any information
relative to the momentum of the produced particles:
\begin{equation}
{\cal G}[u,\bar{u}]\equiv
\sum_{m,n=0}^\infty
\frac{u^m \bar{u}^n}{m!n!}
\int \prod_{i=1}^m \rmd^3\p_i
\prod_{j=1}^n \rmd^3\q_j\;
\Big|\mathcal{M}_{m,n}(\{\p_i\},\{\q_i\})\Big|^2\; .
\end{equation}
Note that this is also equal to:
\begin{equation}
{\cal G}[u,\bar{u}]=
\sum_{m,n=0}^\infty
u^m \bar{u}^n\,P_{m,n}\; ,
\label{eq:G-2nddef}
\end{equation}
where $P_{m,n}$ is the probability to have exactly $m$ particles and
$n$ antiparticles in the final state.  Obviously, this new generating
function can be obtained from the generating functional
$\mathcal{F}[z,\bar{z}]$ by setting the functions $z(\p)$ and
$\bar{z}(\p)$ to constants respectively equal to $u$ and $\bar{u}$:
\begin{equation}
{\cal G}[u,\bar{u}]=\mathcal{F}[z(\p)=u,\bar{z}(\p)=\bar{u}]\; .
\end{equation}
Thanks to this relationship, one can obtain quantities such as those
defined in \eqs(\ref{eq:N2++int}) as ordinary derivatives of
${\cal G}[u,\bar{u}]$.  For instance,
\begin{equation}
\int \rmd^3\p_1 \rmd^3\p_2\; \frac{\rmd N_2^{++}}{\rmd^3\p_1 \rmd^3\p_2}
=
\frac{\partial {\cal G}[u,\bar{u}]}{\partial u^2}\biggr|_{u,\bar{u}=1}
=
\sum_{m,n=0}^\infty m(m-1)\,P_{m,n}\; .
\end{equation}
The last equality is obtained from \eq(\ref{eq:G-2nddef}), and is the
justification for the right hand side in the first of
\eqs(\ref{eq:N2++int}).

\section{Detailed derivation of the solution}
\label{app:solve}
With the variable change from $x^0$ to $\xi$, simple algebraic
procedures lead to $\partial_0=2\omega\xi(1-\xi)\partial_\xi$ and
$\partial_0^2=4\omega^2\xi(1-\xi)[\xi(1-\xi)\partial_\xi^2
 +(1-2\xi)\partial_\xi]$, from which we can rewrite the equation of
motion~\nr{eq:keom} in the following form;
\begin{equation}
 \Bigl[ \xi(1-\xi)\partial_\xi^2 + (1-2\xi)\partial_\xi
  + \mu^2 \xi^{-1} + \nu^2 (1-\xi)^{-1} -\lambda^2 \Bigr]
  \psi_\k(\xi) = 0 \;.
\label{eq:diff}
\end{equation}
In what follows we will explain how to solve this differential
equation.  Before finding the analytical solution, from this form of
the equation we can already confirm that the asymptotic behavior of
the particle solution is
$\rme^{\pm\rmi E_\k^{\textrm{in}} x^0}$ at $x^0\to-\infty$ and
$\rme^{\pm\rmi E_\k^{\textrm{out}} x^0}$ at $x^0\to+\infty$ as it
should be.  We shall pick up the most singular terms out of the
differential equation (\ref{eq:diff}), which gives
\begin{equation}
 \bigl[ \xi\partial_\xi^2 + \partial_\xi + \mu^2 \xi^{-1} \bigr]
  \psi_\k^{\textrm{in}}(\xi) = 0 \;,
\end{equation}
around $\xi=0$ (i.e.\ $x^0\to-\infty$).  It is easy to find the
solution of this equation as
$\psi_\k^{\textrm{in}}(\xi) = \xi^{\pm\rmi\mu}
 \simeq \rme^{\pm\rmi E_\k^{\textrm{in}}x^0}$.  In the same way we can
extract the behavior around $\xi\to 1$ (i.e.\ $x^0\to +\infty$) from
the singular terms;
\begin{equation}
 \bigl[ (1-\xi)\partial_\xi^2 - \partial_\xi
  + \nu^2 (1-\xi)^{-1} \bigr] \psi_\k^{\textrm{out}}(\xi) = 0 \;,
\end{equation}
leading to
$\psi_\k^{\textrm{out}}(\xi) = (1-\xi)^{\pm\rmi\nu}
 \simeq \rme^{\mp\rmi E_\k^{\textrm{out}} x^0}$.

Now let us return to solving \eq(\ref{eq:diff}).  Because we have seen
the boundary condition, it is convenient to factorize the plane-wave
pieces as follows;
\begin{equation}
 \psi_\k(\xi) = \xi^{-\rmi\mu} (1-\xi)^{-\rmi\nu} \varphi_\k(\xi) \;,
\end{equation}
then we can find the equation that $\varphi_\k(\xi)$ should satisfy as
\begin{equation}
 \begin{split}
 & \Bigl\{ \xi(1-\xi)\partial_\xi^2 + \bigl[ 1-2\rmi\mu
  - (-2\rmi\mu-2\rmi\nu+2)\xi \bigr]\partial_\xi \\
 &\qquad\qquad  -(-\rmi\mu-\rmi\nu-\rmi\lambda'
  +1/2)(-\rmi\mu-\rmi\nu+\rmi\lambda'+1/2) \Bigr\} \varphi_\k(\xi) = 0 \;.
 \end{split}
\end{equation}
Here we recall that the hypergeometric differential equation,
\begin{equation}
 \Bigl\{ x(1-x)\partial_x^2 + \bigl[ c-(a+b+1)x \bigr]\partial_x
  -ab \Bigr\} f(x) = 0 \,,
\end{equation}
has two independent solutions given by
\begin{equation}
 f^{(1)}(x) = \Hyg{a}{b}{c}{x} \,,\quad
 f^{(2)}(x) = x^{1-c} \Hyg{a\!+\!1\!-\!c}{b\!+\!1\!-\!c}{2\!-\!c}{x} \;.
\end{equation}
Therefore, by the identification of
\begin{equation}
 a = \frac{1}{2} - \rmi(\lambda'+\mu+\nu) \,,\quad
 b = \frac{1}{2} + \rmi(\lambda'-\mu-\nu) \,,\quad
 c=1-2\rmi\mu \;,
\end{equation}
we finally arrive at the solution (\ref{eq:hyper}).

\bibliographystyle{h-physrev4mod2}
\bibliography{spires}

\end{document}